\documentclass{sig-alternate-05-2015}
 \pdfoutput=1

\usepackage[T1]{fontenc}

\usepackage{graphicx}
\usepackage{booktabs}
\usepackage{amsmath}
\usepackage{amssymb}
\usepackage{amsfonts}
\usepackage{epsfig}
\usepackage{graphicx}
\usepackage{subfigure}
\usepackage{url}
\usepackage[bottom]{footmisc}
\usepackage{balance}
\usepackage{algorithmic}
\usepackage{algorithm}
\usepackage{multirow}
\usepackage{xspace}
\usepackage{times}
\usepackage[singlelinecheck=off]{caption}
\DeclareCaptionType{copyrightbox}
\usepackage{pifont}
\newcommand{\xmark}{\ding{55}}%

\newtheorem{theor}{Theorem}

\newtheorem{lem}{Lemma}

\newtheorem{defn}{Definition}

\newtheorem{exam}{Example}
\newtheorem{problem}{Problem}

\newcommand{\spara}[1]{\smallskip\noindent{\bf #1}}

\newcommand{\squishlist}{
 \begin{list}{$\bullet$}
  {  \setlength{\itemsep}{0pt}
     \setlength{\parsep}{3pt}
     \setlength{\topsep}{3pt}
     \setlength{\partopsep}{0pt}
     \setlength{\leftmargin}{2em}
     \setlength{\labelwidth}{1.5em}
     \setlength{\labelsep}{0.5em}
} }
\newcommand{\squishlisttight}{
 \begin{list}{$\bullet$}
  { \setlength{\itemsep}{0pt}
    \setlength{\parsep}{0pt}
    \setlength{\topsep}{0pt}
    \setlength{\partopsep}{0pt}
    \setlength{\leftmargin}{2em}
    \setlength{\labelwidth}{1.5em}
    \setlength{\labelsep}{0.5em}
} }

\newcommand{\squishdesc}{
 \begin{list}{}
  {  \setlength{\itemsep}{0pt}
     \setlength{\parsep}{3pt}
     \setlength{\topsep}{3pt}
     \setlength{\partopsep}{0pt}
     \setlength{\leftmargin}{1em}
     \setlength{\labelwidth}{1.5em}
     \setlength{\labelsep}{0.5em}
} }

\newcommand{\squishend}{
  \end{list}
}

\newcommand{\eat}[1]{}

\newcommand{\NP}{\ensuremath{\mathbf{NP}}\xspace}
\newcommand{\sharpP}{\ensuremath{\mathbf{\#P}}\xspace}

\newcounter{ccc}

\newcommand{\bigO}{\mathcal{O}}

\begin{document}

\setcopyright{acmcopyright}




\acmPrice{\$15.00}

%

\title{Select Your Questions Wisely: For Entity Resolution \\ With Crowd Errors}

\numberofauthors{3} 

\author{
%
%
\alignauthor
Vijaya Krishna Yalavarthi\\
       \affaddr{NTU Singapore}\\
       \email{yalavarthi@ntu.edu.sg}
\alignauthor
Xiangyu Ke\\
       \affaddr{NTU Singapore}\\
       \email{xiangyu001@e.ntu.edu.sg}
\alignauthor
Arijit Khan\\
       \affaddr{NTU Singapore}\\
       \email{arijit.khan@ntu.edu.sg}
}

\maketitle
\sloppy
\pagenumbering{arabic}

\vspace{-4mm}
\begin{abstract}
Crowdsourcing is becoming increasingly important in entity resolution tasks due to their inherent complexity
such as clustering of images and natural language processing. Humans can provide more insightful information for these
difficult problems compared to machine-based automatic techniques. Nevertheless, human workers
can make mistakes due to lack of domain expertise or seriousness, ambiguity, or even due to malicious intents.
The bulk of literature usually deals with human errors via majority voting or  by assigning a universal
error rate over crowd workers. However, such approaches are incomplete, and often inconsistent,
because the expertise of crowd workers are diverse with possible biases, thereby making it
largely inappropriate to assume a universal error rate for all workers over all crowdsourcing tasks.

We mitigate the above challenges by considering an uncertain graph model, where the edge probability
between two records $A$ and $B$ denotes the ratio of crowd workers who voted YES on the question if $A$ and $B$ are
same entity. To reflect independence across different crowdsourcing tasks,
we apply the notion of possible worlds, and develop
parameter-free algorithms for both next crowdsourcing and entity resolution tasks.
In particular, for next crowdsourcing, we identify the record pair that maximally increases the reliability of the current clustering.
Since reliability takes into account the connected-ness inside and across all clusters, this metric is more effective in deciding
next questions, in comparison with state-of-the-art works, which consider local features, such as individual edges, paths, or nodes to select
next crowdsourcing questions. Based on detailed empirical analysis over real-world datasets, we find that our proposed solution,
{\sf PERC} (\underline{p}robabilistic \underline{e}ntity \underline{r}esolution with imperfect \underline{c}rowd)
improves the quality by 15\% and reduces the overall cost by 50\% for the crowdsourcing-based entity resolution.
\end{abstract}

\vspace{-1mm}
\section{Introduction}
\label{introduction}

Entity Resolution (ER) is the task of disambiguating manifestations of real-world entities in
various records by linking and clustering \cite{GM13}. For example, there could be different
ways of addressing the same person in text, or several photos of a particular object.
Also known as Deduplication, this is a critical step in data cleaning and analytics, knowledge base construction,
comparison shopping, health care, and law enforcement, among many others.

Although machine-based techniques exist for ER tasks, past studies have shown that crowdsourcing can
produce higher quality results, especially for more complex jobs such as classification and clustering
of images, video tagging, optical character recognition, and natural language processing \cite{GWKP11}. Various crowdsourcing services,
e.g., Amazon's Mechanical Turk (AMT) and CrowdFlower \cite{MP15}, 
allow individuals and commercial organizations to set up tasks that humans can perform for certain rewards.
Since a crowd tasker does not work for free, bulk of the literature in this domain
aims at minimizing the cost of crowdsourcing, while also maximizing the overall ER result quality \cite{CLLDF16,WLKFF13,VBD14,WLG13}.
However, human workers can be error-prone due to lack of domain expertise, individual biases, task complexity
and ambiguity, or simply because of tiredness, and malicious behavior \cite{GKRW12,VG15}. As an example, even considering
answers from workers with high-accuracy statistics in AMT, we find that the average crowd error rate can be up to 25\% (we define
average crowd error rate in Section~\ref{sec:experiments}). State-of-the-art works
elude this severe concern by majority voting \cite{WLG13,VBD14,WLKFF13}, that is, to ask the same question to multiple people and
consider the majority answer; or by assigning a universal error rate for crowd taskers \cite{VG15}. Many other works bypass this as an orthogonal
problem to crowdsourced ER, because there are various approaches to compute and reduce crowdsourcing
biases and errors, including \cite{DS09,LLOSWZ12,KOS11}.

\vspace{-1mm}
\spara{Challenges.}
Considering the quality assurance as an orthogonal problem to crowdsourced ER, however, is a substandard solution. Instead,
approaching both these problems together improves the quality of ER, which is evident from recent works \cite{GKRW12,VG15,WXL15}.
The majority voting is often unreliable because spammers and low-paid workers may collude to produce incorrect answers \cite{LLOSWZ12}.
Besides, the tasker crowd is large, anonymous, transient, and it is usually difficult to establish a trust relationship with a
specific worker \cite{KOS11}. Each batch of tasks is solved by a group of taskers who may be completely new, and one may not see them again,
thereby making it unrealistic to assign a universal error rate for all workers over all crowdsourcing tasks.

The major contribution of our work is to develop an end-to-end pipeline for the crowdsourcing-based ER problem, taking into consideration
potential crowd errors. While crowdsourcing a few questions might be sufficient for an initial clustering of records
(e.g., one may crowdsource only $n-1$ record pairs so to construct a spanning tree with all $n$ records),
in order to improve the ER quality, specifically in the presence of crowd errors, crowdsourcing of
more record pairs is necessary. Perhaps, asking the crowd about all $\bigO(n^2)$ record pairs
would provide a very good ER accuracy, but that is prohibitively expensive. Hence, the critical
question that we investigate in this work is as follows.
{\em Given the current clustering, what is the best record pair to crowdsource next}?
Our objective is two-fold: The set of next crowdsourcing questions should be selected in a way that increases
the ER accuracy as much as as possible, at the expenses of as few next crowdsourcing questions as possible.
\begin{table}[tb!]
	\vspace{-2mm}
	\tiny
	\begin{center}
		\begin{tabular} {c||c||cccc||ccc}
			\multicolumn{1}{c||}{\emph{datasets}} & \multicolumn{1}{c||}{\emph{accuracy:}}   & \multicolumn{4}{c||}{\emph{\# crowdsourced questions}}& \multicolumn{3}{c}{\emph{\% crowdsourcing cost}} \\
			\multicolumn{1}{c||}{}                & \multicolumn{1}{c||}{\emph{F1-}}        & \multicolumn{1}{c}{\emph{\sf MinMax}}  & \multicolumn{1}{c}{\emph{\sf DENSE}} & \multicolumn{1}{c}{\emph{\sf PC-Pivot}} & \multicolumn{1}{c||}{\emph{\sf PERC}} & \multicolumn{3}{c}{\emph{reduction by {\sf PERC} over}} \\
			\multicolumn{1}{c||}{}                & \multicolumn{1}{c||}{\emph{measure}} & \multicolumn{1}{c}{\cite{GKRW12}}  & \multicolumn{1}{c}{\cite{VG15}} & \multicolumn{1}{c}{\cite{WXL15}} & \multicolumn{1}{c||}{[this work]} & \multicolumn{1}{c}{\emph{\sf MinMax}}  & \multicolumn{1}{c}{\emph{\sf DENSE}} & \multicolumn{1}{c}{\emph{\sf PC-Pivot}}\\
			\hline \hline
			Allsports	& 0.9 & 13.6K & 16.0K & 21.7K & {\bf 11.7K} & 13.97\% & 26.87\% & 46.08\%\\
			Gymnastics  & 0.9 & 1.3K  & 1.5K  & 1.8K & {\bf 0.8K}  & 38.46\% & 46.67\% & 55.56\%\\
			Landmarks   & 0.9 & 11.0K & 8.0K  & 16K & {\bf 5.9K}  & 46.36\% & 26.25\% & 63.12\%\\
			Cora        & 0.8 & 22.5K & 14.0K & \xmark & {\bf 7.2K} & 68.00\% & 48.57\% & \xmark \\
			\hline \hline
		\end{tabular}
		\caption{\small Crowdsourcing cost reduction by {\sf PERC}: We present the number of crowdsourcing questions required to achieve a certain accuracy
			for various methods. A detailed description about our datasets, accuracy measure, and experiment setting can be found in Section~\ref{sec:experiments}. \label{tab:summary}}
		\vspace{1mm}
	\end{center}
	\vspace{-8mm}
\end{table}

Given its practical importance, not surprisingly, the problem of identifying the next question for crowdsourced ER, in the presence of crowd errors,
has been studied recently: {\sf MinMax}\cite{GKRW12}, {\sf PC-Pivot} \cite{WXL15}, and {\sf DENSE} \cite{VG15}.
{\em These methods consider ad-hoc, local features to select next questions, such as
	individual paths (e.g., {\sf MinMax}), nodes (e.g., {\sf PC-Pivot}), or the set of either positive or negative edges (e.g., {\sf DENSE, shown in Appendix}).
	Hence, they generally fail to capture the strength of the entire clustering, resulting in higher crowdsourcing cost to achieve a reasonable ER accuracy}.

\spara{Our Contribution.}
As opposed to local metrics used in prior works, we select the next crowdsourcing question by
considering the strength of the entire clustering. Our global metric, denoted as the {\em reliability},
follows the notion of connected-ness in an uncertain graph.
Intuitively, reliability measures how well-connected a cluster is, and
also how well-separated two clusters are. We then
systematically identify the next crowdsourcing question, either from a weakly connected cluster,
or across a pair of clusters that are weakly separated, thereby creating a balance between
stronger and weaker components in the clustering. As a consequence, our reliability-based
next crowdsourcing algorithm reduces the crowdsourcing cost significantly, which is evident
in Table~\ref{tab:summary}.

Our contributions can be summarized as follows.
\begin{itemize}
	\setlength\itemsep{0.01em}
	\item For the next crowdsourcing problem, we introduce a novel metric called ``reliability'' of a clustering, that measures connected -ness
	within and across clusters by following the notion of uncertain graphs (Section~\ref{sec:crowd}).
	This is more effective than local-feature-based next crowdsourcing approaches
	\cite{GKRW12,WXL15,VG15}, as demonstrated with our running example (Section~\ref{sec:crowd}) and also verified in our experimental results (see Table~\ref{tab:summary}).
	\item Using reliability-based next crowdsourcing, we develop an end-to-end solution, {\sf PERC}, for crowdsourced ER (Section~\ref{sec:er}).
	Our algorithms are parameter-free in the sense that we do not require any user-defined threshold values, and no apriori information about the error rate of the crowd workers.
	\item We perform detailed experiments with four real-world datasets using Amazon's Mechanical Turk platform. The performance analysis illustrates the
	quality, cost, and efficiency improvements of our framework (Section~\ref{sec:experiments}).
\end{itemize}
\begin{figure}[tb!]
	\centering
	\includegraphics[scale=0.24]{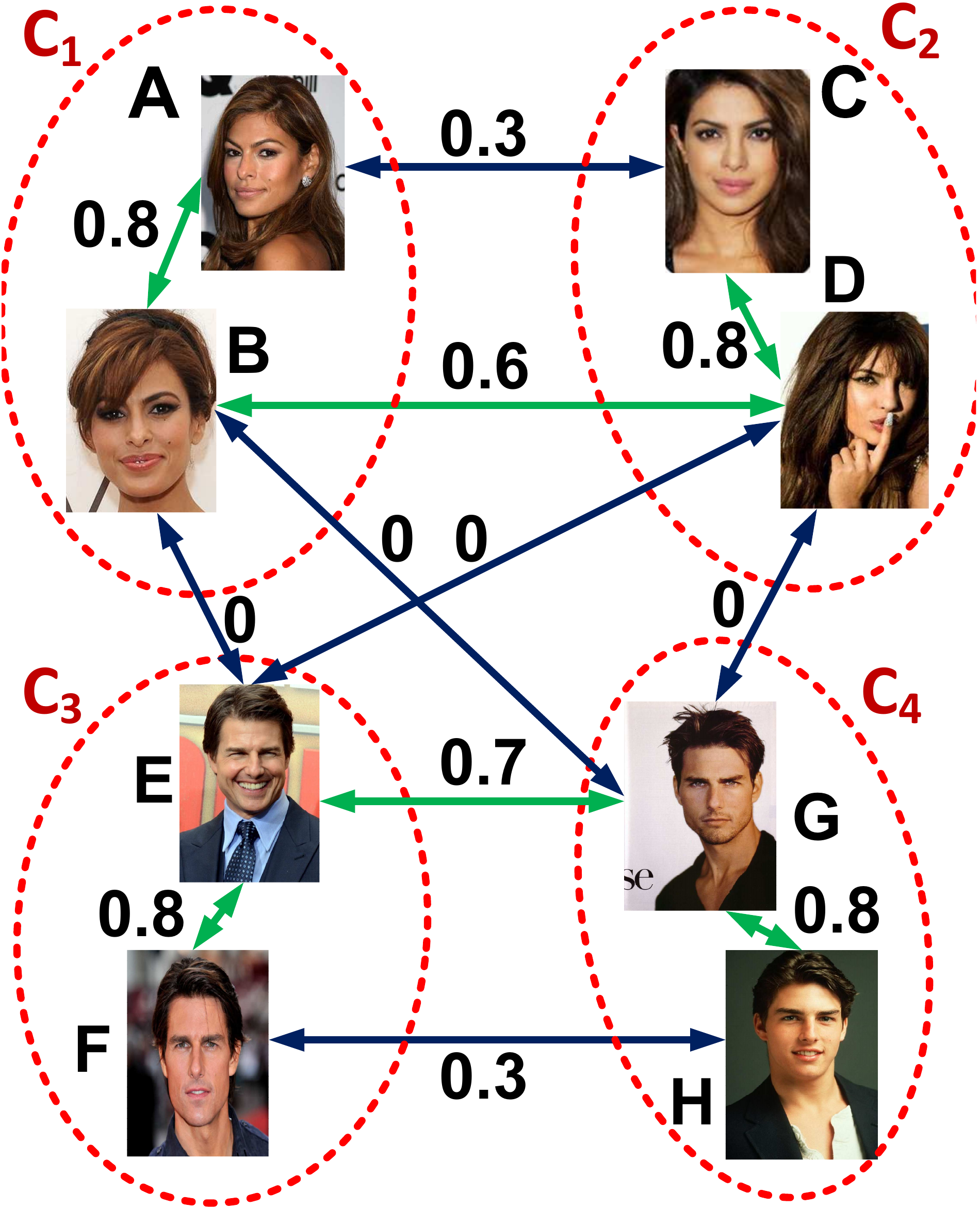}
	\vspace{-3mm}
	\caption{\small Running example: Edge probability
		denotes ratio of crowd workers voted YES for the respective records pair to be same entity.}
	\label{fig:motivation_universal}
	\vspace{-5mm}
\end{figure}

\spara{Running Example.}
Consider a dataset of eight images shown in Figure~\ref{fig:motivation_universal}. Records $A$, $B$ belong to famous American actress and model, {\em Eva Mendes}; $C$, $D$ to Bollywood star and lead actress of the American television series, Quantico,
{\em Priyanka Chopra}; and $E$, $F$, $G$, $H$ to Hollywood actor {\em Tom Cruise}.
80\% of crowd workers voted YES that both records in each of the following pairs are same:
$\langle A,B \rangle$, $\langle C,D \rangle$, $\langle E,F \rangle$, and $\langle G,H \rangle$.
All crowd workers also answered NO for the edges between the following
cluster pairs: $\langle \mathbb{C}_1, \mathbb{C}_3\rangle$, $\langle \mathbb{C}_2, \mathbb{C}_3\rangle$, $\langle \mathbb{C}_1, \mathbb{C}_4\rangle$, and $\langle \mathbb{C}_2, \mathbb{C}_4\rangle$. In this example, four clusters $\mathbb{C}_1, \mathbb{C}_2, \mathbb{C}_3$ and $\mathbb{C}_4$ are formed, as shown in Figure~\ref{fig:motivation_universal}. Our objective is to identify the next question to crowdsource that maximizes the gain. It can be observed that asking a question between clusters $\mathbb{C}_3$ and $\mathbb{C}_4$ is more beneficial because all images in $\mathbb{C}_3$ and $\mathbb{C}_4$ belong to the same entity, and one more edge with probability greater than 0.5 helps in merging these two clusters.

\vspace{-1mm}
\section{Preliminaries}
\label{sec:problem_formulation}

\subsection{Background}

\spara{Entity Resolution (ER).}
An ER algorithm receives an input set of records $R=\{r_1,r_2,\ldots,r_n\}$ and a pairwise
similarity function $F$, and it returns a set of matching pair of records: $\mathbb{C}=\{R_1, R_2,\ldots, R_m\}$, such that,
$R_i \cap R_j = \phi$ for all $i,j$, and $\cup_{i} R_i = R$. We call each $R_i$ a {\em cluster} of $R$, and
each cluster represents a distinct real-world entity. The partition of $R$ into a set of clusters is
called a {\em clustering} $\mathbb{C}$ of $R$. If $r_1$ and $r_2$ are matching (non-matching), they are denoted by $r_1=r_2$ ($r_1 \ne r_2$).

An ER algorithm generally obeys the two following relations.

\noindent \underline{Transitivity.} Given three records $r_1, r_2,$ and $r_3$, if $r_1=r_2$ and $r_2=r_3$, then we have $r_1=r_3$.

\noindent \underline{Anti-transitivity.} Given three records $r_1, r_2,$ and $r_3$, if $r_1=r_2$ and $r_2\ne r_3$, then we have $r_1\ne r_3$.

Thus, a clustering $\mathbb{C}$ of the input set $R$ of records is transitively closed. One can derive the following theorem combinatorially.
We omit the proof due to limitation of space.
\begin{theor}
\label{th:cluster_numbers}
For $n$ records, there can be $(2^n-n)$ different clusterings, where each cluster in some clustering must have between $(1,n)$ records.
\end{theor}

\vspace{-1mm}
\spara{Crowdsourced ER.}
We use a crowdsourcing platform such as Amazon's Mechanical Turk (AMT), which provides APIs for conveniently
using a large number of human workers to complete micro-tasks (also known as Human Intelligent Tasks (HITs)). To identify
whether two records belong to the same entity, we create an HIT for the pair, and publish it to AMT
with possible binary answers: A worker needs to submit `YES' if she thinks that the record pair is matching, and `NO' otherwise.

For mitigating crowd errors, we allow multiple workers to perform the same HIT. We then assign an edge with probability $p(r_i,r_j)$
between two records $r_i$ and $r_j$, where $p(r_i,r_j) \in (0,1)$ denotes the ratio of crowd workers who voted YES on the question if $r_i$ and $r_j$ are
same entity.

\spara{Uncertain Graph.}
Every HIT creates an uncertain, undirected edge between the respective record pair, thereby generating an uncertain, undirected graph $\mathcal{G}=(R,E,p)$,
as depicted previously in Figure~\ref{fig:motivation_universal}. Each record $r_i\in R$ denotes a node in the graph, $E\subseteq R\times R$ represents the set
of edges between the record pairs that were crowdsourced, and $p(e) \in (0,1)$ is the probability of the edge $e \in E$ as derived earlier.
In our context, it is important to note that $p(e)=0$ (i.e., all crowd workers voted non-matching) is {\em not}
equivalent to the edge $e$ being absent in $\mathcal{G}$ (i.e., the pair is not crowdsourced yet).

To reflect independence across different crowdsourcing tasks (i.e., each HIT can be performed by a different set of workers),
we employ the well-established notion of possible world, together with the assumption that each edge can be matching or non-matching,
independent of other edges \cite{JLDW11}.  Hence, the uncertain graph $\mathcal{G}$ yields $2^{|E|}$ deterministic graphs (or, possible worlds) $G \sqsubseteq \mathcal{G}$,
where each $G$ is a pair $(R,E_G)$, with $E_G \subseteq E$ are matching edges, and its probability of being observed is given in Equation~\ref{eq:possible_world}.
\begin{align}
\displaystyle P(G) = \prod_{e\in E_G}p(e) \prod_{e\in E\setminus E_G}(1 - p(e))
\label{eq:possible_world}
\end{align}

Next, we have the following observation.
\vspace{-1mm}
\begin{lem}
Every clustering of the input record set $R$ corresponds to some possible world of the uncertain graph $\mathcal{G}=(R,E,p)$. However,
every possible world of $\mathcal{G}$ might not be a clustering of $R$.
\end{lem}

\begin{figure}[t!]
\centering
\includegraphics[scale=0.23]{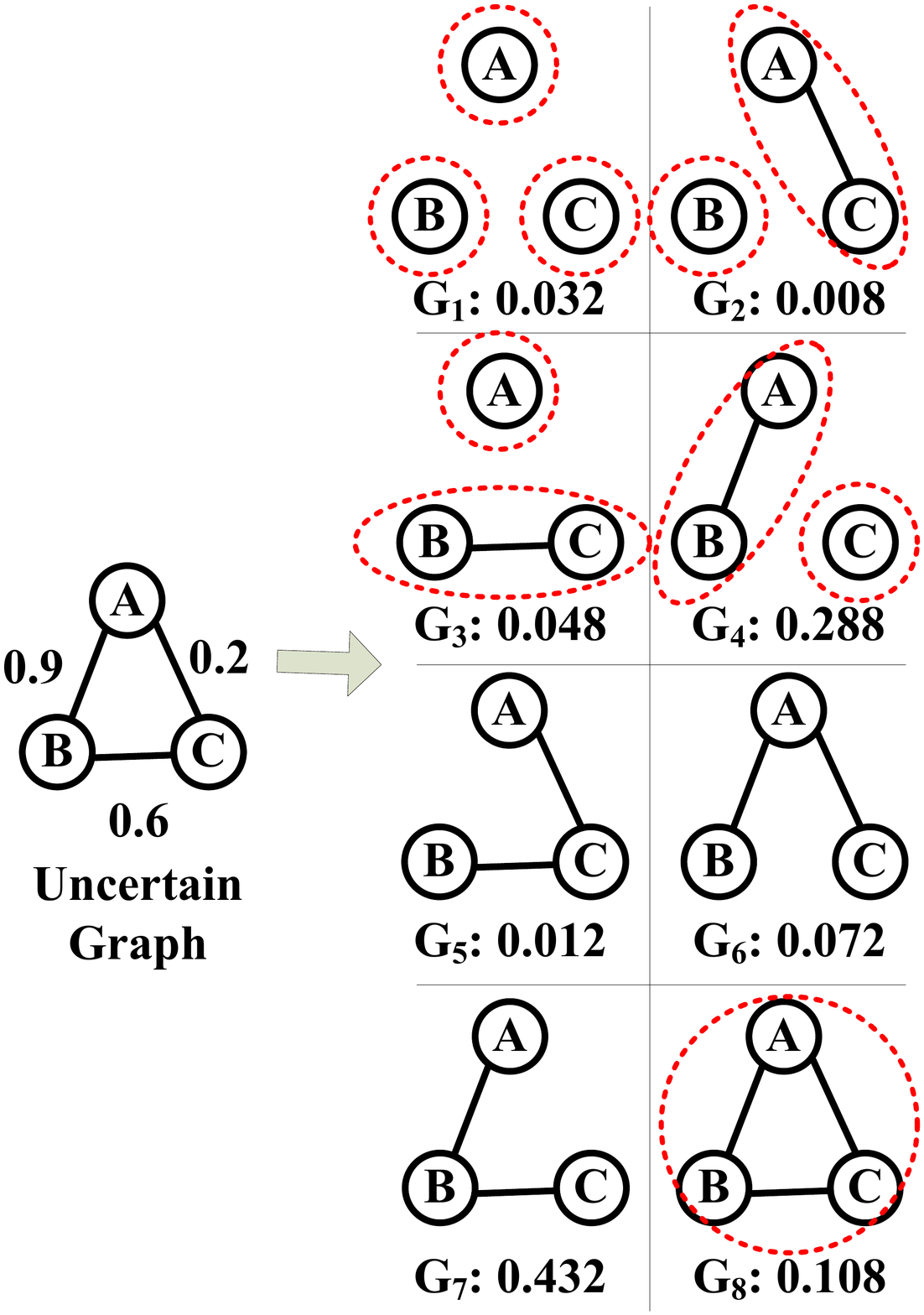}
\vspace{-3mm}
\caption{\small Possible worlds of an uncertain graph: Three possible worlds $G_5$, $G_6$, $G_7$ are not clusterings, as they
are not transitively closed. For example, in $G_5$,  $A=C$ and $C=B$, but $A\ne B$, thus violating transitivity.}
\label{fig:possible_world}
\vspace{-5mm}
\end{figure}

The first part of the lemma is trivial (i.e., follows from the definition of a possible world), whereas the second part
holds since every possible world is not transitively closed. We demonstrate this fact with an example in Figure~\ref{fig:possible_world},
where three possible worlds $G_5,G_6,$ and $G_7$ of the given uncertain graph are not clusterings.

Since every clustering corresponds to some possible world, we define the likelihood of a clustering as the probability of the respective possible
world being observed. In Figure~\ref{fig:possible_world}, the likelihood of the clustering $\{(A,B),(C)\}$ is same as $P(G_4)$, which is $0.288$.

\vspace{-1mm}
\subsection{Entity Resolution Problem}

Given $R,\mathcal{G}$, let us consider a clustering $\mathbb{C}=\{R_1,R_2,\ldots,R_m\}$ of $R$. We define the likelihood
of $\mathbb{C}$ as the probability that (1) all edges inside every cluster $R_i$ exist, and (2) all edges across every pair
of clusters $R_j,R_k$ do not exist. Since an edge can exist independent of others, we compute the  likelihood $L(\mathbb{C})$
as follows.
\begin{align}
& L(\mathbb{C})=\prod_{R_i \in \mathbb{C}} \left[\prod_{e\in E \cap (R_i \times R_i)}p(e)\right] \times \prod_{\substack{{R_j,R_k \in \mathbb{C}}\\{j<k}}} \left[\prod_{e\in E \cap (R_j \times R_k)}\left(1- p(e)\right)\right] &
\end{align}

We formally introduce the ER problem below.
\vspace{-1mm}
\begin{problem}[Entity Resolution]
\label{prob:mlc}
Given the set $R$ of records and an uncertain graph $\mathcal{G}=(R,E,p)$, find the (transitively closed) clustering $\mathbb{C}$ of $R$
having the highest likelihood $L(\mathbb{C})$.
\end{problem}

\vspace{-1mm}
The problem of finding the most-likely clustering (also referred to as the maximum-likelihood clustering), however, is \NP-hard, which can be verified by a polynomial-time reduction from the
\NP-hard correlation clustering problem \cite{VG15}.
\vspace{-1mm}
\begin{theor}
\label{th:np}
Given an uncertain graph $\mathcal{G}=(R,E,p)$ over records set $R$, finding the maximum-likelihood clustering of $R$ is \NP-hard.
\end{theor}

\vspace{-1mm}
Correlation clustering is the most natural setting for clustering a set of records that are connected by both positive and negative edges \cite{HCLM09}.
Many approximate and heuristic algorithms were proposed for correlation clustering \cite{ES09,VG15}. Indeed, all prior works such as \cite{GKRW12,VG15,WXL15}
in the domain of crowdsourced ER, that incorporated human error, also employed correlation clustering. Therefore, in our {\sf PERC} framework,
we apply correlation clustering for the ER problem. Details about our clustering algorithm will be given in Section~\ref{sec:er}.
We shall first introduce our next crowdsourcing algorithm in the following, which is the key contribution of this work.

\vspace{-1mm}
\section{ Next Crowdsourcing}
\label{sec:crowd}

We discuss our algorithm for selecting the next crowdsourcing question. We assume that
an initial (maximum-likelihood) clustering $\mathbb{C}$ is already constructed from the records set $R$
and the uncertain graph $\mathcal{G}=(R,E,p)$, and now we want to identify the best
entity pair $\langle r_i, r_j \rangle \not\in E$ to crowdsource next.

\vspace{-1mm}
\subsection{Reliability of a Clustering}

Intuitively, our objective is to identify a pair $\langle r_i, r_j \rangle \not\in E$ that can improve the quality
of the given clustering as much as possible. To this end, we identify the two following ``connected-ness''-based criteria
that determine the quality of a clustering $\mathbb{C}$. Let us denote $\mathbb{C}=\{R_1,R_2, . . . ,R_m\}$,
where each $R_i$ is a cluster and represents a distinct real-world entity.
\begin{itemize}
	\setlength\itemsep{0.01em}
	\item How well each cluster $R_i$ is connected?
	\item How well every pair of clusters $R_j,R_k$ $(j<k)$ is disconnected?
\end{itemize}
\begin{figure}[t!]
	\centering
	\includegraphics[scale=0.21]{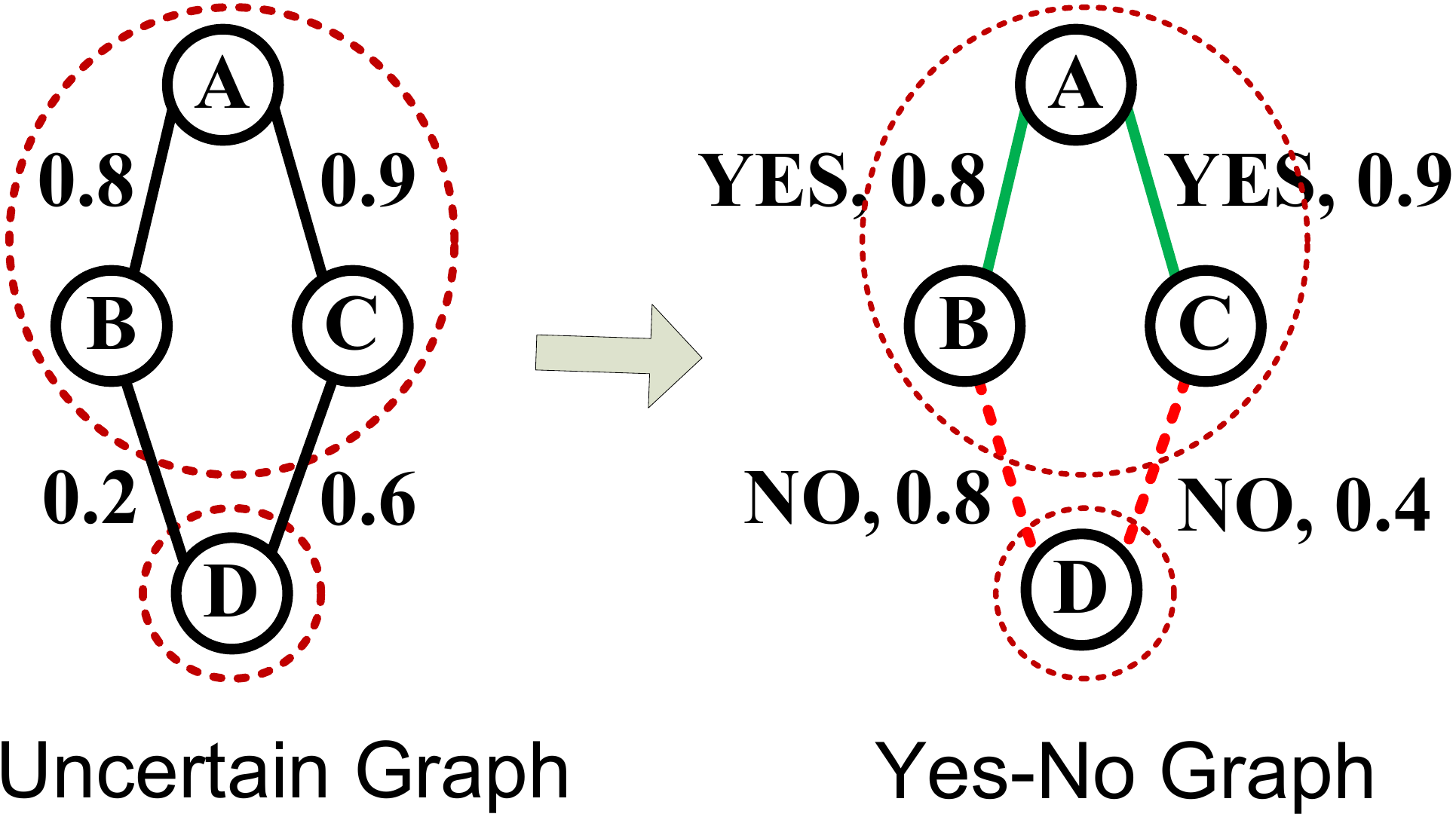}
	\vspace{-3mm}
	\caption{\small Reliability of a clustering}
	\label{fig:reliability}
	\vspace{-5mm}
\end{figure}

Given a clustering $\mathbb{C}=\{R_1,R_2, . . . ,R_m\}$ and the uncertain graph $\mathcal{G}=(R,E,p)$, all edges inside a cluster are called YES edges, whereas
the edges across two clusters are referred to as NO edges. If $e\in E$ is an YES edge, we define its existence probability $p_Y(e)=p(e)$.
On the other hand, if $e \in E$ is a NO edge, we compute its existence probability as $p_N(e)=1-p(e)$.
We derive an YES-NO graph $\mathcal{G}_{Y|N}=(R,E,p_{Y|N},L)$ from the uncertain graph $\mathcal{G}$ as follows.
$\mathcal{G}_{Y|N}$ has the same set of nodes and edges as $\mathcal{G}$, but each edge $e$ in $\mathcal{G}_{Y|N}$
has a binary label $L(e)$, which can be either YES or No, as defined above. For a YES edge $e$, its probability $p_{Y|N}(e)=p_Y(e)$.
For a NO edge $e$, its probability $p_{Y|N}(e)=p_N(e)$. Next, we formalize the notion of connectivity and disconnectivity.
\vspace{-1mm}
\begin{defn}[Connectivity]
	Given a cluster $R_i$ and the YES-NO graph $\mathcal{G}_{Y|N}$, the connectivity of $R_i$ is defined as the sum of the probability of
	those possible worlds of $\mathcal{G}_{Y|N}$ where all records in $R_i$ are connected by YES edges.  Formally,
	\begin{align}
		Connect(R_i) = \sum_{G \sqsubseteq \mathcal{G}_{Y|N}} [I(R_i,G) \times P(G)]
		\vspace{-3mm}
	\end{align}
	\vspace{-2mm}
\end{defn}
\vspace{-2mm}
In the above equation, $I(R_i,G)$ is an indicator function over a possible deterministic graph $G \sqsubseteq \mathcal{G}_{Y|N}$
taking value $1$ if records in $R_i$ are all connected (by YES edges) in $G$, and $0$ otherwise.
\vspace{-1mm}
\begin{defn}[Disconnectivity]
	Given a pair of clusters $R_j, R_k$ $(j<k)$ and the YES-NO graph $\mathcal{G}_{Y|N}$, the disconnectivity between $R_j,R_k$ is defined as the sum of the probability of
	those possible worlds of $\mathcal{G}_{Y|N}$ where at least one NO edge exists between $R_j$ and $R_k$. Formally,
\vspace{-1mm}
	\begin{align}
		& Disconnect(R_j,R_k)  & \\
		&= \begin{cases}
			0                 \qquad \qquad \qquad  \qquad \qquad \qquad \qquad     ;\text{if } (R_j \times R_k)\cap E = \phi \nonumber & \\
			1-\prod_{(r_i,r_l) \in (R_j \times R_k)\cap E}(1-p_N(r_i,r_l)) \quad \qquad ;\text{otherwise}
		\end{cases}
		\vspace{-3mm}
	\end{align}
	\vspace{-2mm}
\end{defn}
Based on the above definition, we observe that for all $i,j,k$; $j<k$, the following events are independent. (1) A cluster $R_i$ is connected, and (2) a
pair of clusters $R_j,R_k$ are disconnected. Therefore, one can multiply the probability of these events to measure the overall quality of a clustering $\mathbb{C}$.
For practical reasons, we avoid multiplying fractions, and instead compute summation over logarithms (Equation~\ref{equ:reliability}). Thus, if either
of $Connect(R_i)$ or $Disconnect(R_j,R_k)$ is zero, we substitute it by a very small positive fraction. Formally, we denote this metric as the {\em reliability} of a clustering.
\begin{defn}[Reliability]
	Given a clustering $\mathbb{C}=\{R_1,R_2,$ $\ldots,R_m\}$ and the YES-NO graph $\mathcal{G}_{Y|N}$, the reliability of $\mathbb{C}$ is defined as the probability that
	every cluster $R_i$ is connected and every pair of clusters $R_j,R_k$ $(j<k)$ is disconnected, i.e.,
	\begin{align}
		Rel(\mathbb{C}) = \sum_i \log \left(Connect\left(R_i\right)\right) + \sum_{j<k}\log \left(Disconnect\left(R_j,R_k\right)\right)
		\label{equ:reliability}
		\vspace{-3mm}
	\end{align}
	\vspace{-2mm}
\end{defn}
\vspace{-3mm}
\begin{exam}
	In Figure ~\ref{fig:reliability}, we compute the reliability of the clustering ${\mathbb C}=\{(A,B,C),(D)\}$. We first construct the YES-NO graph on the right. Then, we have: $Connect(A,B,C)=0.72$,
	$Connect(D)=1.0$, and $Disconnect\left(\left(A,B,C\right),\left(D\right)\right)=1-(1-0.8)(1-0.4)=0.88$. Hence, $Rel(\mathbb{C})=\log 0.72+\log 1+\log 0.88 \approx -0.20$.
\end{exam}

\vspace{-2mm}
\subsection{Next Crowdsourcing Problem}

We derive, for every record pair $\langle r_i,r_j\rangle \not \in E$, the improvement in reliability of the already computed clustering $\mathbb{C}$,
if one crowdsources the pair, and thereby assigns the edge probability $p(r_i,r_j)$. However, one does not know $p(r_i,r_j)$
apriori. Therefore, we consider an optimistic scenario, that is, for all possible values of $p(r_i,r_j)\in(0,1)$, we derive what will be the maximum
possible increment in $Rel(\mathbb{C})$ by crowdsourcing  $\langle r_i,r_j\rangle$. We select the record pair that maximally increases $Rel(\mathbb{C})$,
under such optimistic assumption.

Our formulation has several desirable features, such as monotonicity and improving weaker components, as stated next.
\begin{lem}
	For any new edge $e$ that we crowdsourced, $Rel(\mathbb{C})$ will increase maximally when $p_{Y|N}(e)=1$.
\end{lem}
In other words, if the new edge $e$ is inside a cluster (i.e., YES edge),
then its probability requires to be $p(e)=1$, which means that all workers agreed on the record pair as matching. On the other hand, if the new edge $e$ is across two clusters (i.e., NO edge),
then its probability must be $p(e)=0$, which implies that all workers agreed on the record pair as non-matching. To put it simply, if the next crowdsourcing result is fully consistent
with our previous clustering, then the quality of the clustering improves maximally.

\begin{lem}
	By adding a new edge $e$, the reliability of $\mathbb{C}$ remains the same when $p_{Y|N}(e)=0$. It increases monotonically as we have larger values of $p_{Y|N}(e)$.
\end{lem}
Generally speaking, the more is the ratio of workers who agree with the previous clustering, the higher is the improvement in the clustering quality.
\begin{lem}
	For any new edge $e$ that we included, if $p_{Y|N}(e)> 0.5$, the maximum-likelihood clustering, as defined in Problem~\ref{prob:mlc}, remains the same for the updated graph.
	\label{lem:no_change}
\end{lem}

This implies that if the majority of the crowd workers agree with our previous clustering, there is no need to change the clustering.

Below, we formally introduce the next crowdsourcing problem.
\begin{problem}[Next Crowdsourcing]
	Given the set $R$ of records, an uncertain graph $\mathcal{G}=(R,E,p)$, and
	a clustering ${\mathbb C}$, find the record pair $\langle r_i, r_j \rangle \not\in E$,
	such that adding an edge $(r_i,r_j)$, with $p_{Y|N}(r_i,r_j)=1$, maximally increases the reliability of $\mathbb{C}$.
	\label{prob:next_crowd}
	\vspace{-1mm}
\end{problem}

\vspace{-2mm}
\subsection{Demonstration with Running Example}
\begin{figure}[tb!]
	\centering
	\includegraphics[scale=0.17]{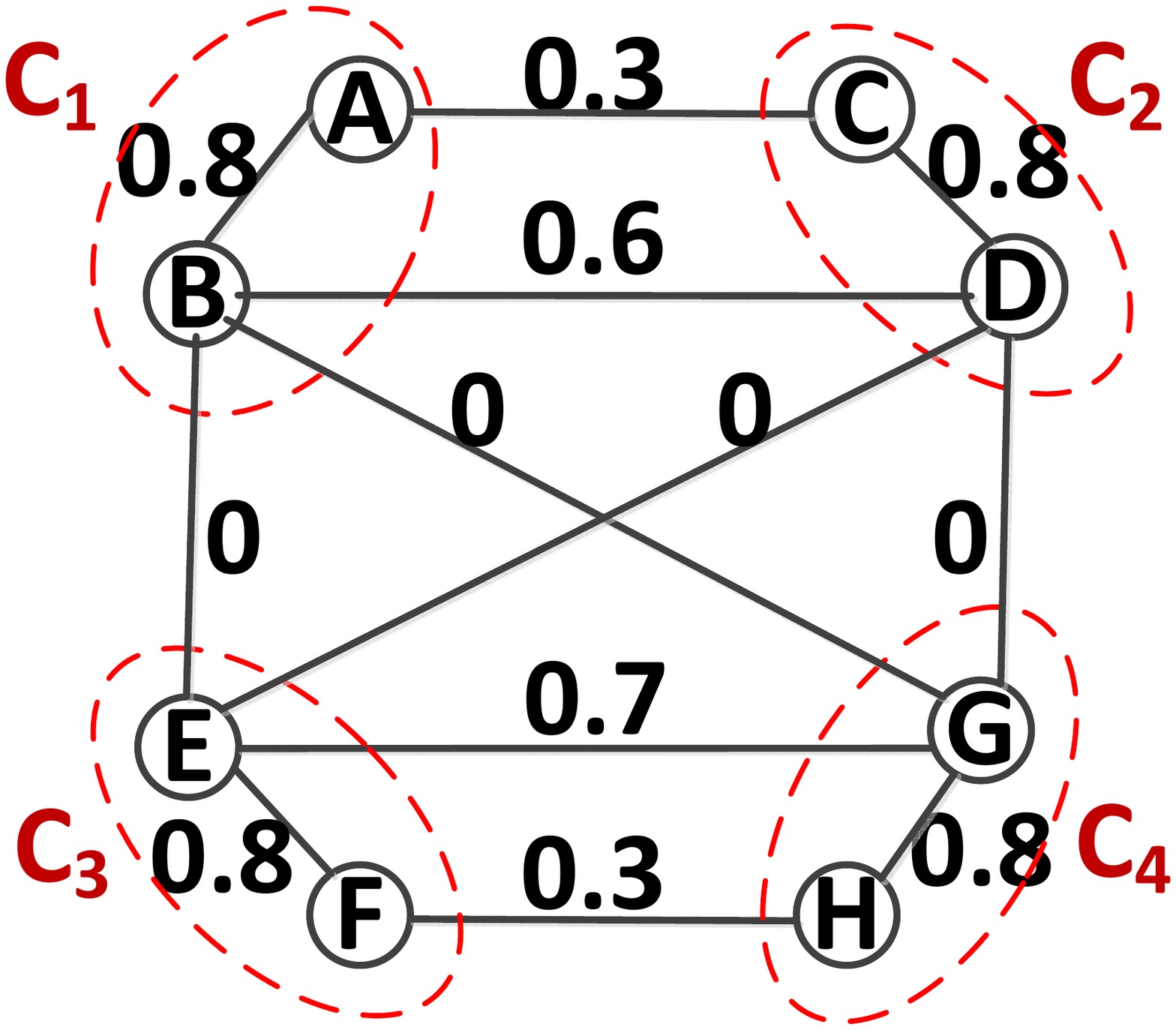}
	\vspace{-12mm}
	\caption{\small Reliability-based next crowdsourcing: running example}
	\label{fig:running_2}
	\vspace{-5mm}
\end{figure}

We now demonstrate how our reliability-based next crowdsourcing technique deals with the running example in Figure \ref{fig:motivation_universal}.
\begin{exam}
	\vspace{-1mm}
	Figure~\ref{fig:running_2} is the abstract version of our running example in Figure~\ref{fig:motivation_universal}.
	The clustering algorithm identifies four clusters: $\mathbb{C}_1=\{A,B\}$, $\mathbb{C}_2=\{C,D\}$, $\mathbb{C}_3=\{E,F\}$,
	and $\mathbb{C}_4=\{G,H\}$. Each cluster has connectivity 0.8. The disconnectivity values across these clusters are as follows.
	$Disconnect(\mathbb{C}_1, \mathbb{C}_2)$ = 0.82, $Disconnect(\mathbb{C}_2,$ $\mathbb{C}_3)$ = 1,
	$Disconnect(\mathbb{C}_1,\mathbb{C}_3)$ = 1, and $Disconnect(\mathbb{C}_3,\mathbb{C}_4)$ = 0.79.
	As $Disconnect(\mathbb{C}_3, \mathbb{C}_4)$ is the least among all others, our algorithm priorities crowdsourcing
	an edge across $\mathbb{C}_3, \mathbb{C}_4$. Intuitively, the separation between $\mathbb{C}_3, \mathbb{C}_4$ is the weakest, thus we require to
	ask more questions about this separation. The reliability gain by adding a new edge $e$ between $\mathbb{C}_3, \mathbb{C}_4$,
	having probability $p_{Y|N}(e)=1$, is $\log1-\log0.79$=0.10; whereas,
	the reliability gain by adding a new edge $e$ between $\mathbb{C}_1, \mathbb{C}_2$,
	with probability $p_{Y|N}(e)=1$, is $\log1-\log0.82$=0.08. Hence, for next crowdsourcing, our algorithm
	selects an edge across $\mathbb{C}_3, \mathbb{C}_4$. Indeed,
	one more edge with probability greater than 0.5 across $\mathbb{C}_3, \mathbb{C}_4$
	helps in merging these two clusters, while an edge with probability less than 0.5 will make their separation stronger.
	This is consistent with our running example that asking a question across clusters $\mathbb{C}_3$ and $\mathbb{C}_4$ is more beneficial.
\end{exam}

\vspace{-1mm}
\spara{Remarks.} As demonstrated with the running example, our next crowdsourcing method
usually prioritizes the weaker components and improves their quality, thereby creating a balance between the quality
of stronger and weaker components in the clustering. This is evident if we consider two pairs of clusters
such that $Disconnect(R_1,$ $R_2) < Disconnect(R_3,R_4)$, then our method will always prioritize a pair $\langle r_1,r_2 \rangle
\in R_1 \times R_2$ over any other pair $\langle r_3,r_4\rangle \in R_3 \times R_4$, for the next crowdsourcing.
For brevity, let us denote by $d_1=Disconnect(R_1,$ $R_2)$ and $d_2=Disconnect(R_3,R_4)$.
In the first case, we consider an edge $(r_1,r_2)$ with $p_N(r_1,r_2)=1$, i.e., $p(r_1,r_2)=0$. Hence, the increase in reliability,
following Equation~\ref{equ:reliability},
is $\log(1/d_1)$. Analogously, in the second case, the increase in reliability
is $\log(1/d_2)$. Since $d_1<d_2$, the pair $\langle r_1,r_2 \rangle$
is preferred over $\langle r_3,r_4\rangle$.

In case of connectivity of individual clusters, in general no such relationship exists. However, if the connectivity
of one cluster is significantly smaller than that of the other, e.g.,
$Connect(R_1) << Connect(R_2)$, it is very likely that our method will select a pair from $R_1$ for the next crowdsourcing problem.
Let $c_1=Connect(R_1)$ and $c_2=Connect(R_2)$. Also, assume that $\delta_1$ is the maximum increase in
$c_1$ if we add an edge $e$ of probability $p_Y(e)=1$ (i.e., $p(e)=1$) in $R_1$. Similarly, let $\delta_2$ be the maximum increase in
$c_2$ if we add an edge $e'$ of probability $p_Y(e')=1$ (i.e., $p(e')=1$) in $R_2$. Hence, in the first case, the increase in reliability is $\log (1+\delta_1/c_1)$,
whereas in the second case, the increase in reliability is $\log (1+\delta_2/c_2)$. Since $c_1<<c_2$, it is very likely that
$\delta_1/c_1 > \delta_2/c_2$. Therefore, in such cases,
our method will prioritize a specific record pair from $R_1$ over all pairs from $R_2$, for the next crowdsourcing problem.

\vspace{-1mm}
\subsection{Next Crowdsourcing Algorithm}
\vspace{-1mm}
\spara{Difficulties.}
A na\"{\i}ve algorithm to find the best record pair for next crowdsourcing would be inefficient due to the following challenges.
\begin{itemize}
	\setlength\itemsep{0.01em}
	\item Computing the connectivity of a cluster, also known as the {\em all-terminal-reliability problem in device networks},
	is \sharpP-hard \cite{JLDW11}. Hence, finding the exact connectivity value, even for a modest size cluster, is almost infeasible.
	\item At each round of crowdsourcing, we identify the best record pair not in $E$. Usually, the uncertain graph $\mathcal{G}$
	is sparse, that is, $|E| << \bigO(|R|^2)$. Therefore, at every round, one needs to compare $\bigO(|R|^2)$ pairs in order to identify the best one
	for next crowdsourcing.
\end{itemize}

\vspace{-1mm}
\spara{Monte Carlo Sampling.}
Due to its intrinsic hardness, we tackle the connectivity estimation problem from an approximation viewpoint. We use the answer computed by
Monte Carlo (MC) sampling as a proxy. This is a reasonable choice as MC-sampling is an unbiased estimator, thus by running it for a sufficiently
large number of times, its answer is expected to converge to the real answer with a high probability.  In particular, we first sample $t$
possible graphs, $G_1, G_2, \ldots, G_t$ of a subgraph of $\mathcal{G}_{Y|N}$ induced by the nodes in some cluster $R_i$,
according to (YES) edge probability $p_{Y|N}=p_{Y}$. We then compute the ratio of possible graphs which are connected,
out of $t$ possible graphs that were generated. This gives the MC-estimation of connectivity for cluster $R_i$.
To speed up the sampling process, we combine MC-sampling with a breadth first search (BFS) from one of the
nodes in $R_i$ \cite{JLDW11}. If the maximum numbers of nodes and edges in a cluster are $n_{max}$ and $e_{max}$, respectively,
then the time complexity of MC-based connectivity estimation is given by $\bigO\left(t\left(n_{max}+e_{max}\right)\right)$.
Based on empirical results over our datasets, we observed that the MC-estimator converges with a number
of samples $t \approx 1000$. This is roughly the same number observed in \cite{JLDW11} for MC-sampling based
reliability estimation over other real-world uncertain graphs.
\begin{algorithm}[tb!]
	\caption{Next Crowdsourcing Algorithm}
	\small
	\begin{algorithmic}[1]\label{alg:next}
		\REQUIRE Records set $R$, uncertain graph $\mathcal{G} = (R,E,p)$, clustering $\mathbb{C}$
		\ENSURE Record pair $\langle r_i, r_j \rangle \not \in E$ to be crowdsourced next
		\STATE Let $\mathbb{C}=\{R_1,R_2,\ldots,R_m\}$
		\IF {Clustering updated last round}
		\STATE priority queue $Q = \phi$
		\FORALL{$R_i$}
		\FORALL{$\langle r_j,r_k \rangle \in (R_i\times R_i)\setminus E$}
		\STATE Form $\mathcal{G}'$ by adding $(r_j,r_k)$ in $\mathcal{G}$, with $p_Y(r_j,r_k)=1$
		\STATE $prio(r_j,r_k)= Rel_{\mathcal{G}'}(\mathbb{C})-Rel_{\mathcal{G}}(\mathbb{C})$
		\STATE Insert $\left(\langle r_j,r_k \rangle, prio(r_j,r_k)\right)$ into $Q$
		\ENDFOR
		\ENDFOR
		\FORALL{$(R_j\times R_k)$, $j<k$}
		\STATE Find one $\langle r_i,r_l \rangle \in (R_j\times R_k)\setminus E$
		\STATE Form $\mathcal{G}'$ by adding $(r_i,r_l)$ in $\mathcal{G}$, with $p_N(r_i,r_l)=1$
		\STATE $prio(r_i,r_l)= Rel_{\mathcal{G}'}(\mathbb{C})-Rel_{\mathcal{G}}(\mathbb{C})$
		\STATE Insert $\left(\langle r_i,r_l \rangle, prio(r_i,r_l)\right)$ into $Q$
		\ENDFOR
		
		/* Clustering not Updated in last round  */
		
		\ELSE
		\IF{last edge was inserted in $R_i$}
		\FORALL{$\langle r_j,r_k \rangle \in (R_i\times R_i)\setminus E$}
		\STATE Form $\mathcal{G}'$ by adding $(r_j,r_k)$ in $\mathcal{G}$, with $p_Y(r_j,r_k)=1$
		\STATE $prio(r_j,r_k)= Rel_{\mathcal{G}'}(\mathbb{C})-Rel_{\mathcal{G}}(\mathbb{C})$
		\STATE Update $\left(\langle r_j,r_k \rangle, prio(r_j,r_k)\right)$ into $Q$
		\ENDFOR
		\ENDIF
		\IF{last edge was inserted between $R_j$ and $R_k$, $j<k$}
		\STATE Find one $\langle r_i,r_l \rangle \in (R_j\times R_k)\setminus E$
		\STATE Form $\mathcal{G}'$ by adding $(r_i,r_l)$ in $\mathcal{G}$, with $p_N(r_i,r_l)=1$
		\STATE $prio(r_i,r_l)= Rel_{\mathcal{G}'}(\mathbb{C})-Rel_{\mathcal{G}}(\mathbb{C})$
		\STATE Insert $\left(\langle r_i,r_l \rangle, prio(r_i,r_l)\right)$ into $Q$
		\ENDIF
		\ENDIF
		\STATE $\langle r_i,r_j \rangle$ = $Q.pop()$
		\STATE return $\langle r_i,r_j \rangle$
	\end{algorithmic}
	\vspace{-1mm}
\end{algorithm}

\vspace{-1mm}
\spara{Algorithm.}
The complete method for next crowdsourcing is given in Algorithm~\ref{alg:next}.
Let us denote by {\em priority} of a pair $\langle r_i,r_j\rangle \not \in E$ as the increase in reliability of the existing clustering,
when the edge $(r_i,r_j)$ is included with probability  $p_{Y|N}(r_i,r_j)=1$ (lines 7 and 14, Algorithm~\ref{alg:next}).
At every round, we crowdsource the record pair with the highest priority. However, priority computation
for all pairs at every round would be expensive. We discuss below how
one can minimize the required number of priority computations.

We note that for a specific round, the priority of all the following record pairs $\langle r_k, r_l \rangle \in (R_i\times R_j)\setminus E$,
for a certain $R_i$ and $R_j$, are the same. Therefore, we compute the priority of only one record pair across every pair of clusters
(lines 11-16, Algorithm~\ref{alg:next}). Finally, if
an edge was inserted in some cluster $R_i$ in the last round and there is no change in the previous clustering (lines 18-24,
Algorithm~\ref{alg:next}), then the priority of the
pairs inside other clusters, as well as those across two clusters, will not change. Similarly, if
an edge was inserted between two clusters $R_i, R_j$ in the last round and there is no change in the earlier clustering
(lines 25-30, Algorithm~\ref{alg:next}),
the priority of the pairs inside all clusters, as well as those across other cluster pairs, will not change. All of these reduce
the priority re-computation necessary for at most $\bigO(n^2_{max})$ pairs at every round, if there is no change in the previous clustering.

In reality, $n_{max}$ is small, around 30$\sim$350 records, for the real-world
datasets that we have considered (and also used by state-of-the-art approaches \cite{GKRW12,WXL15,VG15}). Thus,
overall time complexity of our next crowdsourcing algorithm is $\bigO\left(n^2_{max}\left(t\left(n_{max}+e_{max}\right)\right)\right)$.
In fact, the priority of each record pair inside a cluster can be computed in parallel, and/or one may
sample a selected number of record pairs, uniformly at random, from the cluster; thereby further
reducing the time required to select the next crowdsourcing question.

\spara{Asking Next Questions in Batches.} Algorithm~\ref{alg:next} selects a single question to ask next to the crowd workers.
Instead, one may consider a batch version to issue multiple
high-quality questions. For a batch size $k$ ($k$ is a tunable input parameter), we select
the $k$ record pairs having the highest priority. It is expected that by issuing multiple questions in batches,
the overall quality would decrease, because one does not know the corresponding edge probabilities apriori;
and therefore, we compute the priority of a record pair in an optimistic manner. However, asking questions in batches helps in
reducing the running time
of crowdsourced ER, because many crowd workers would be able to answer the questions in a batch
in parallel. 
\begin{figure}[t!]
	\centering
	\includegraphics[scale=0.24]{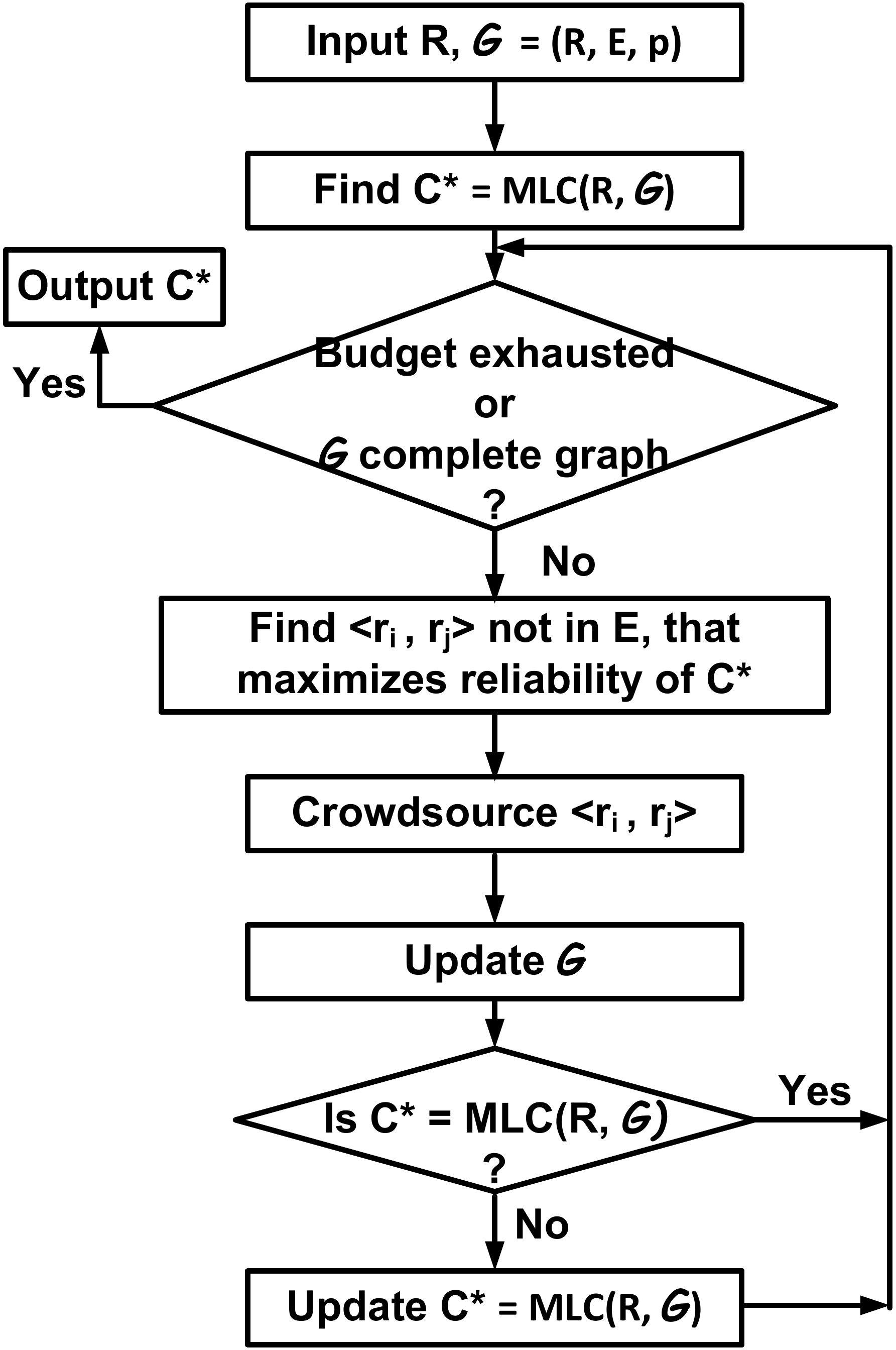}
	\vspace{-3mm}
	\caption{\small Overview of our {\sf PERC} framework}
	\label{fig:workflow}
	\vspace{-5mm}
\end{figure}

\vspace{-2mm}
\section{THE {\sf PERC} FRAMEWORK}
\label{sec:er}

The reliability-based next crowdsourcing method (Section~\ref{sec:crowd}) forms the crux of our {\sf PERC} framework.
Clearly, given a set of records and their similarity values obtained via next crowdsourcing, one requires to
cluster these records. We discuss our clustering technique in Section~\ref{sec:er_clustering},
and then provide in Section~\ref{sec:pipeline} the complete pipeline that combines our next crowdsourcing
and clustering algorithms.

\vspace{-1mm}
\subsection{Clustering Algorithm}
\label{sec:er_clustering}

Given the records set $R$ and an uncertain graph $\mathcal{G} = (R,E,p)$,
we use correlation clustering to find the maximum-likelihood clustering
of $R$ (Problem~\ref{prob:mlc}). We recall that all prior works in crowdsourced ER,
e.g., \cite{GKRW12,VG15,WXL15}, which incorporated human error, also employed correlation clustering.
Since correlation clustering is \NP-hard, several
approximate and heuristic algorithms exist \cite{ES09}. We empirically compare them,
and find the {\em Spectral-Connected-Components} (SCC) technique to be the most effective one.
This is also the same clustering method used in {\sf DENSE} entity resolution \cite{VG15}.

\spara{Spectral-Connected-Components (SCC).} This algorithm starts from the record pair having the highest probability of being the same entity, given the
answers for these two records. If this probability is higher than 0.5, SCC merges the two records into one cluster. In each successive step, the algorithm
finds the clusters with the highest probability of being the same entity, given the answers between them. If this probability is higher than 0.5, the two
clusters are merged into one cluster. Otherwise, SCC stops merging clusters, and returns as output the current set of clusters.

Given two clusters $R_i$ and $R_j$, SCC computes the probability $Pr(R_i,R_j)$ of merging them as given in Equation~\ref{equ:scc}.
\vspace{-0.5mm}
\begin{align}
	&\displaystyle Pr(R_i,R_j) \nonumber & \\
	\vspace{-2mm}
	& = \frac{\displaystyle \prod_{(r_k,r_l) \in (R_i \times R_j)\cap E} p\left(r_k,r_l\right)}{\displaystyle \prod_{(r_k,r_l) \in (R_i \times R_j)\cap E} p\left(r_k,r_l\right) + \displaystyle \prod_{(r_k,r_l) \in (R_i \times R_j) \cap E}\left(1-p\left(r_k,r_l\right)\right)} &
	\label{equ:scc}
\end{align}
\vspace{-2mm}

Let the numbers of nodes in the uncertain graph $\mathcal{G}$ be $n$. Then, the time complexity of SCC clustering is $\bigO(n^2)$.
\vspace{-2mm}
\begin{exam}
	We demonstrate SCC clustering with our running example in Figure~\ref{fig:running_2}.
	The algorithm identifies record pairs containing the maximum edge weight (i.e., 0.8).
	We initially clusters any of $A, B$; $C, D$; $E, F$; or $G, H$.
	Later, we continue to cluster another three record pairs as they have the same maximum edge weight.
	Once the four clusters $\mathbb{C}_1=\{A,B\}, \mathbb{C}_2=\{C,D\}$, ${\mathbb{C}_3}=\{E,F\}$, and $\mathbb{C}_4=\{G,H\}$ are identified,
	we verify the edge weights across these cluster. SCC merges two clusters only if the benefit of merging (Equation~\ref{equ:scc}) is more than 0.5.
	Let us consider the merging of clusters $\mathbb{C}_1$ and $\mathbb{C}_2$. Their probability of merging is $\frac{0.3\times0.6}{0.3\times0.6+(1-0.3)\times(1-0.6)}$
	= 0.39. In fact, none of the cluster pairs qualify for merging, and SCC reports $\mathbb{C}_1, \mathbb{C}_2$, ${\mathbb{C}_3}$, and $\mathbb{C}_4$ as the four clusters.
\end{exam}
\vspace{-2mm}
\subsection{Putting Everything Together}
\label{sec:pipeline}

we provide the entire pipeline of our {\sf PERC} framework in Figure~\ref{fig:workflow}.
Given an input set $R$ of records, and the initial uncertain graph $\mathcal{G}$ (which might have no edges in the beginning, or
only a few edges based on initial crowdsourcing), we find the most-likely clustering (MLC) $\mathbb{C}$ of $R$, with SCC algorithm.
Next, we iteratively find the best record pair $\langle r_i, r_j \rangle$ and crowdsource it, until our budget is exhausted, or we
already find a complete (uncertain) graph over $R$.  After every crowdsourcing task, we add an uncertain edge between the respective
record pair, thereby updating $\mathcal{G}$.
\begin{figure}[tb!]
	\vspace{-2mm}
	\centering
	\includegraphics[scale=0.19]{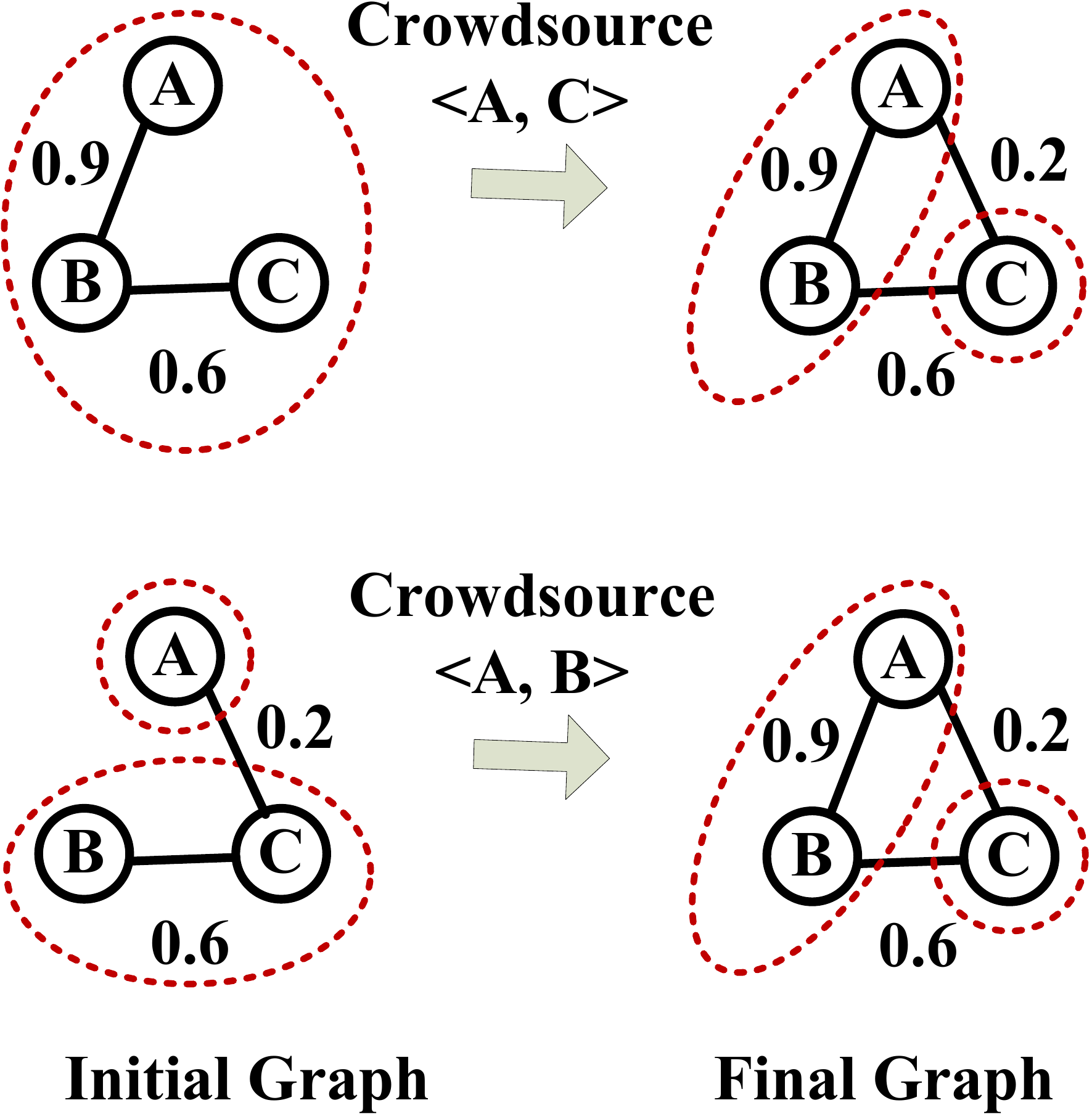}
	\vspace{-3mm}
	\caption{\small Most-likely clustering (MLC) changes due to addition of edges. Above: crowdsourcing result of $\langle A,C\rangle$ changes MLC from $\{(A,B,C)\}$ to $\{(A,B),(C)\}$. Below: crowdsourcing result of $\langle A,B\rangle$ changes MLC from $\{(A),(B,C)\}$ to $\{(A,B),(C)\}$.}
	\label{fig:correction}
	\vspace{-4mm}
\end{figure}

An interesting feature of our framework is that at the end of every round, we
check if the previous MLC $\mathbb{C}$ still remains the MLC for the updated graph.
This can be quickly verified based on Lemma~\ref{lem:no_change}, that is,
if the majority of the crowd workers agree with our previous clustering, there is no need to change the clustering.
Otherwise, we recompute the new MLC and proceed to identify the best record pair to crowdsource for this new MLC.
{\em Such re-clustering enables us to rectify mistakes that might have been incurred at earlier
	rounds due to incomplete information and crowd errors, thereby quickly converging to a high-quality solution}.
We illustrate this feature of our framework with two examples in Figure~\ref{fig:correction}.
As one may observe, in both cases with the additional crowdsourcing evidences,
the new MLC is more promising than the earlier one.

While such updates in the MLC clustering are quite effective, we empirically found that these updates
happen only 20$\sim$25\% of the times after next crowdsourcing. This illustrates that while updating the
previous clustering is critical to improve the ER quality, it does not significantly
impact the total computation time.

\begin{table} [tb!]
\vspace{-1mm}
\tiny
\vspace{-2mm}
\centering
\begin{tabular} { l|cccl }
{\textsf{Dataset}} & {\textsf{\# Records}}  & {\textsf{\# Entities}}          &  {\textsf{\# Record-Pairs}}   &   {\textsf{Crowd Error Rate}}\\
                   &                        &                                 &  {\textsf{Crowdsourced}}   &                              \\ \hline\hline
{\em All Sports}   &     267                &  86                             &            35\,511               &   5.67\% (10 ques. / pair) \\ \hline
 {\em Gymnastics}  &     94                 &  12                             &            4\,371                &   10.65 \% (5 ques. / pair)\\ \hline
 {\em Landmarks}   &     529                &  15                             &            30\,070               &   4.82 \%  (5 ques. / pair) \\ \hline
 {\em Cora}        &     949                &  165                            &            29\,281               &   27.77 \% (5 ques. / pair) \\ \hline
\end{tabular}
\vspace{-2mm}
\caption{\small Properties of datasets}
\label{tab:data}
\vspace{-4mm}
\end{table}

\vspace{-2mm}
\section{Experimental Results}
\label{sec:experiments}

We present empirical results with four real-world, benchmark datasets (three image datasets
and one text dataset). We evaluate entity resolution (ER) accuracy, efficiency, and crowdsourcing
cost of {\sf PERC} under various initial conditions, and by asking the next crowdsourcing
questions one at a time and also in batches, with different crowd errors.
We compare {\sf PERC} with four state-of-the-art crowdsourced ER approaches:
transitive closure ({\sf TC}) clustering \cite{WLG13,VBD14,WLKFF13}, {\sf MinMax} \cite{GKRW12},
{\sf PC-Pivot} \cite{WXL15}, {\sf DENSE} and {\sf bDENSE} \cite{VG15}.
\vspace{-2mm}
\subsection{Environment Setup}

The code is implemented in Python and we perform experiments on a single core of a 32GB, 2.40GHz Xeon server.
All results are averaged over 10 runs. We present our results with spectral-connected-components (SCC)
based correlation clustering, as it performs the best compared other correlation clustering methods \cite{ES09,VG15}.

\vspace{-1mm}
\spara{$\bullet$ Datasets.} We use four benchmark, real-world datasets (Table~\ref{tab:data}) from the literature
of crowdsourced ER.

\vspace{-1mm}
\spara{AllSports:} The {\em AllSports} dataset \cite{VG15} consists of athlete images from different sports, with each image showing a single athlete.

\vspace{-1mm}
\spara{Gymnastics:}  The {\em Gymnastics} dataset \cite{VG15} contains athlete images, but only from gymnastics, and it is more difficult to
distinguish the face of an athlete in this dataset, e.g., the athlete may be upside down on uneven bars.

\vspace{-1mm}
\spara{Landmarks:} The {\em Landmarks} dataset \cite{GKRW12}
has images from 9 cities. We consider a subset of the original dataset, consisting 529 images of
15 different Landmarks.

\vspace{-1mm}
\spara{Cora:} This is a text dataset containing references of scientific publications \cite{WXL15}.
{\em Cora} is one of the largest datasets considered in the literature
of crowdsourced ER, thus we use this dataset for demonstrating scalability.

We use Amazon's Mechanical Turk for crowdsourcing, and follow the same setting that was employed by Verroios and Molina in \cite{VG15},
e.g., considering answers from workers with high-accuracy statistics. We omit the details due to lack of space. In particular,
for {\em AllSports}, we engage 10 workers for each task, whereas 5 workers are employed for each task  in the other datasets \cite{GKRW12,VG15,WXL15}.
For {\em AllSports} and {\em Gymnastics}, due to their smaller sizes, we crowdsource all record pairs. On the contrary,
for {\em Landmarks} and {\em Cora} datasets, we crowdsource about 22\% and 7\%, respectively, of all record pairs, based on next crowdsourcing questions.

All these datasets come with the ground truth clustering results, which we refer to as the {\em gold standard clustering}.
If a worker answered the record pair wrongly, then it is considered an error (Table~\ref{tab:data}). As an example,
out of 10 workers, if 8 workers answered correct and 2 answered wrong, then the error in answering that particular records pair is 20\%.
Crowd error is measured as the average of all such errors over all crowdsourced record pairs.
\begin{figure*}[t!]
\vspace{-2mm}
\centering
\subfigure[\small {\em AllSports}] {
\includegraphics[scale=0.17, angle=270]{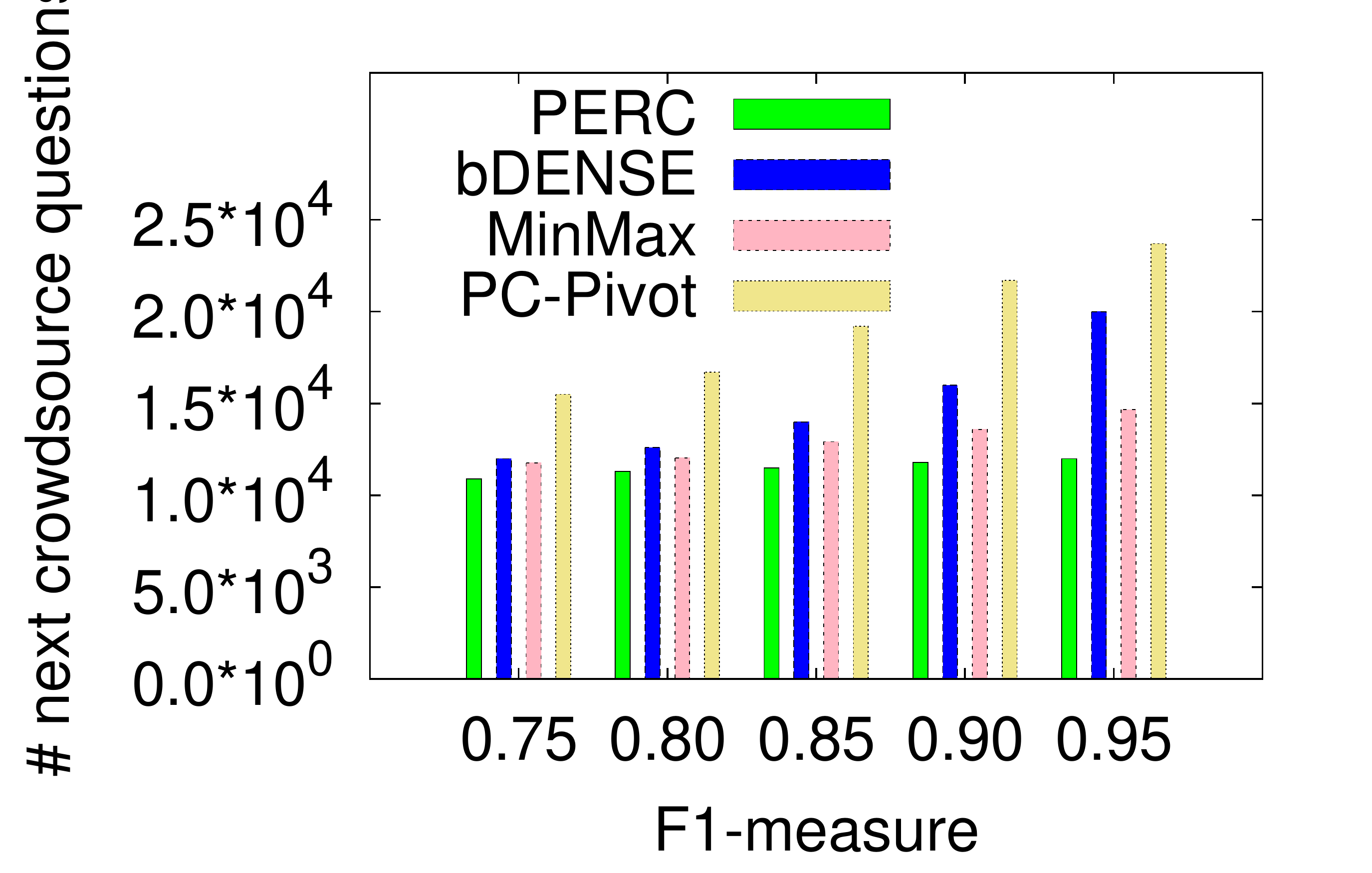}
\label{fig:cost_allsports}
}
\subfigure[\small {\em Gymnastics}]  {
\includegraphics[scale=0.17, angle=270]{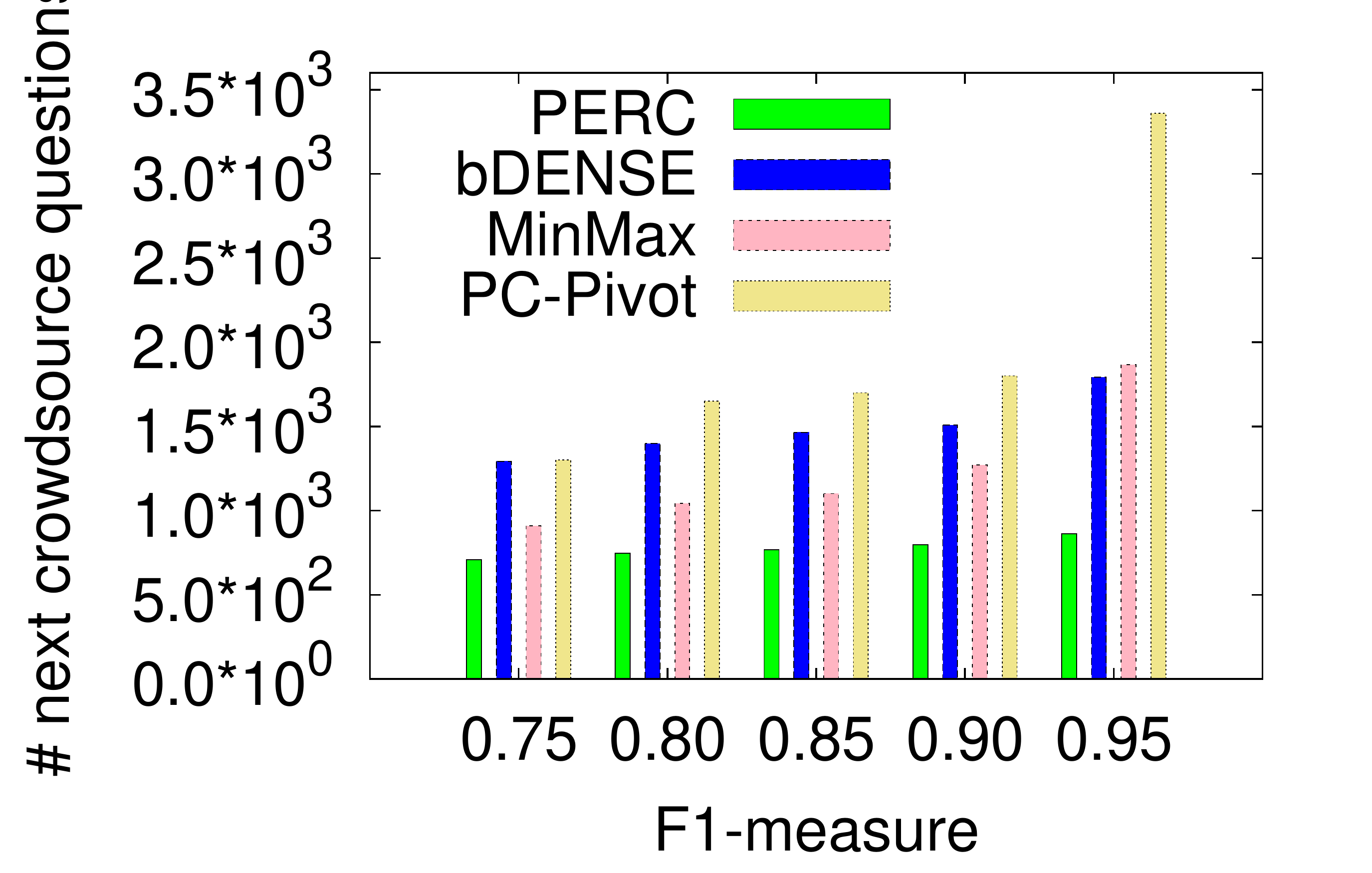}
\label{fig:cost_gym}
}
\subfigure[\small {\em Landmarks}]  {
\includegraphics[scale=0.17, angle=270]{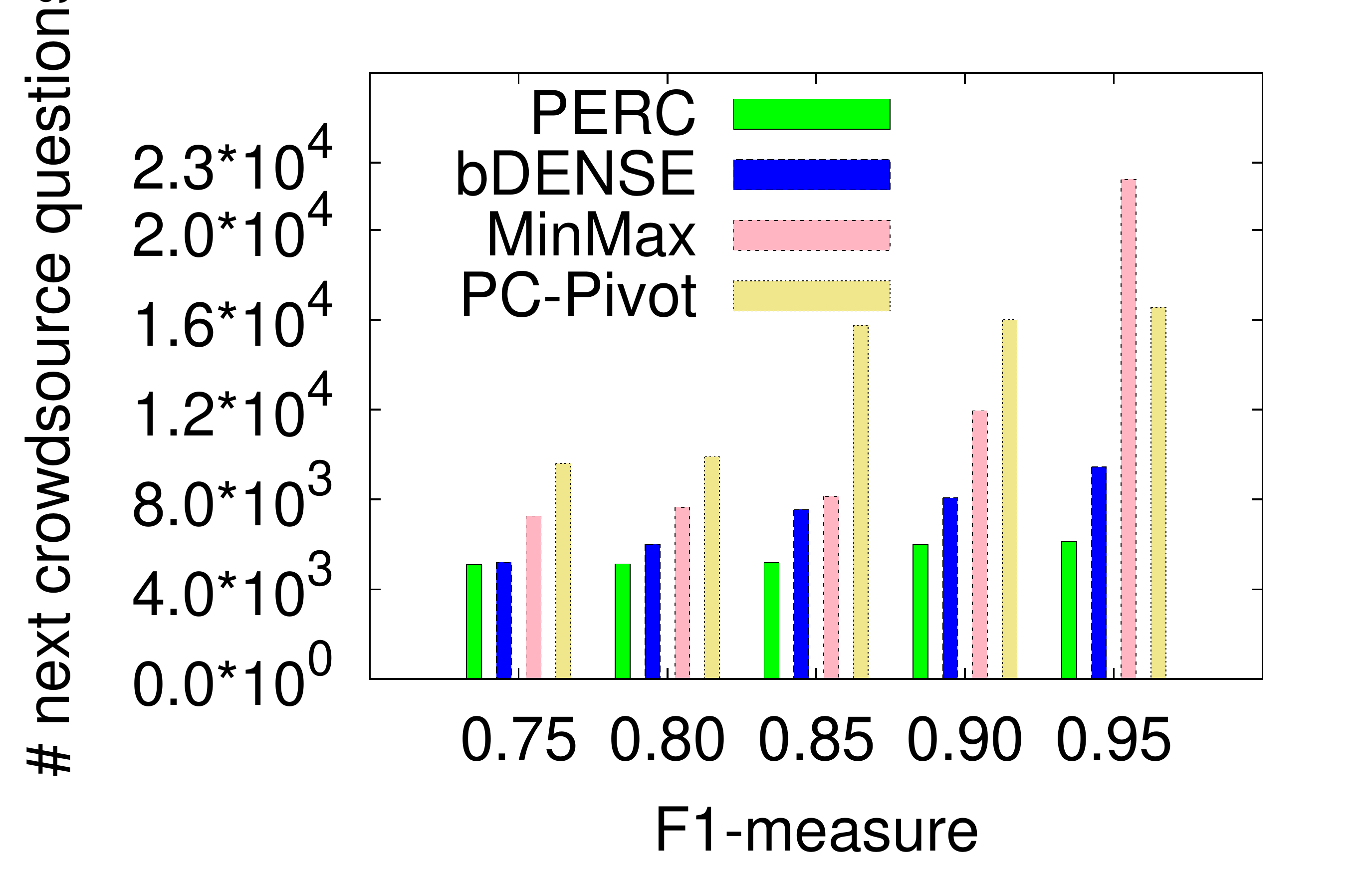}
\label{fig:cost_Landmarks}
}
\vspace{-5mm}
\caption{\small Cost improvement: \# next crowdsourcing questions required to reach a certain accuracy (F1-measure)}
\label{fig:cost_next}
\vspace{-5mm}
\end{figure*}
\begin{figure*}[t!]
\centering
\subfigure[\small {\em AllSports}] {
\includegraphics[scale=0.17, angle=270]{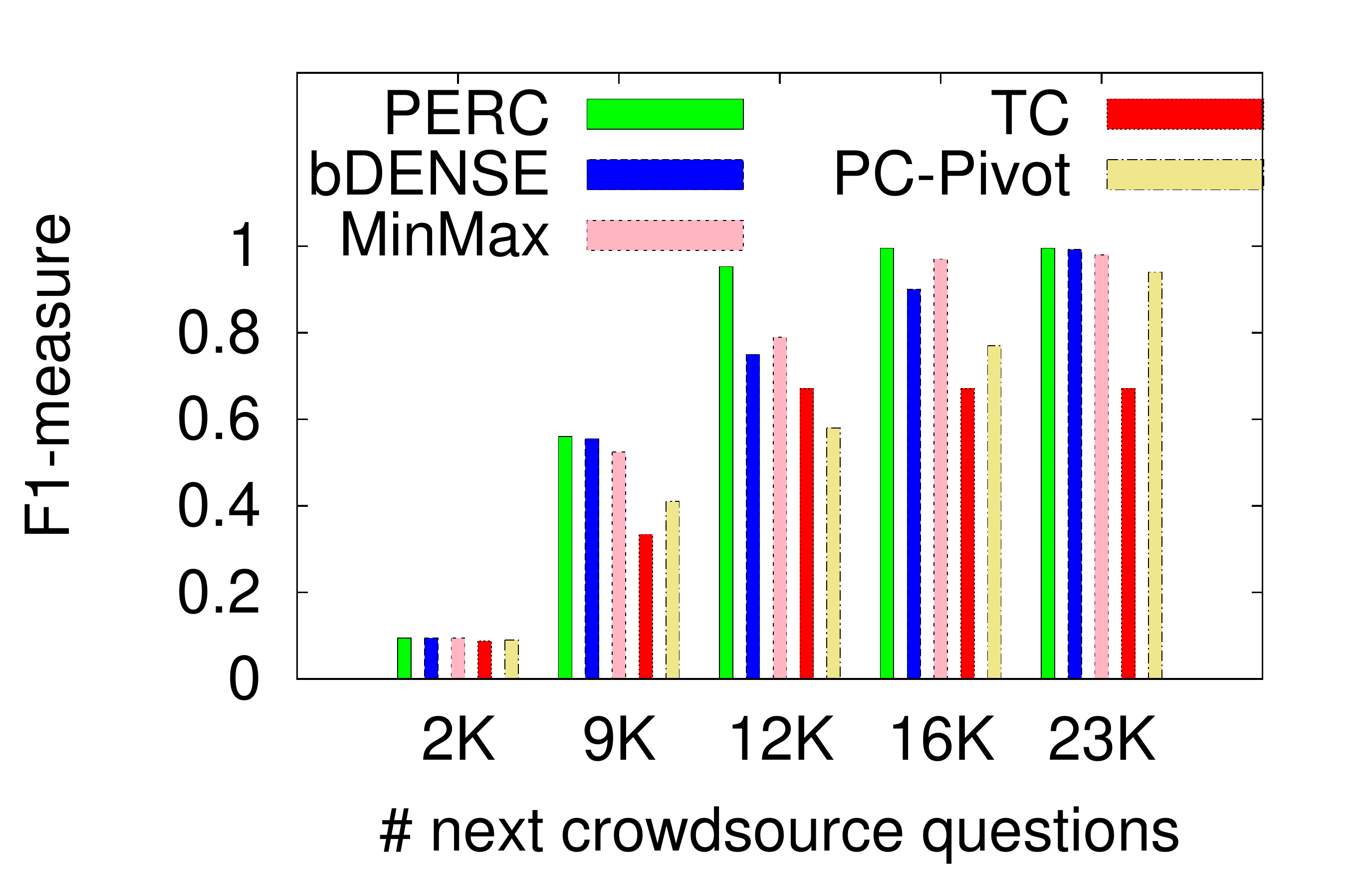}
\label{fig:f1_allsports}
}
\subfigure[\small {\em Gymnastics}]  {
\includegraphics[scale=0.17, angle=270]{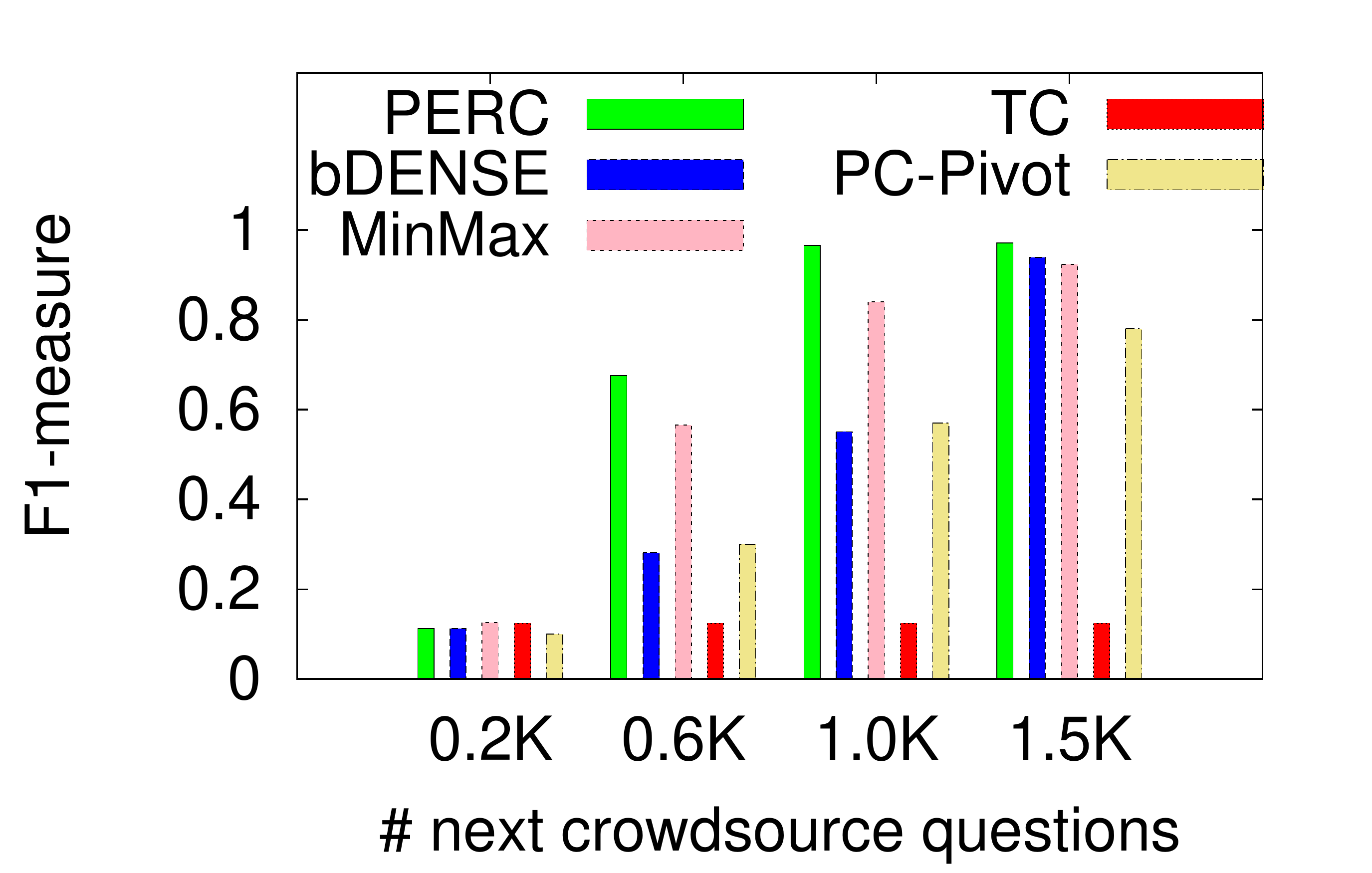}
\label{fig:f1_gym}
}
\subfigure[\small {\em Landmarks}]  {
\includegraphics[scale=0.17, angle=270]{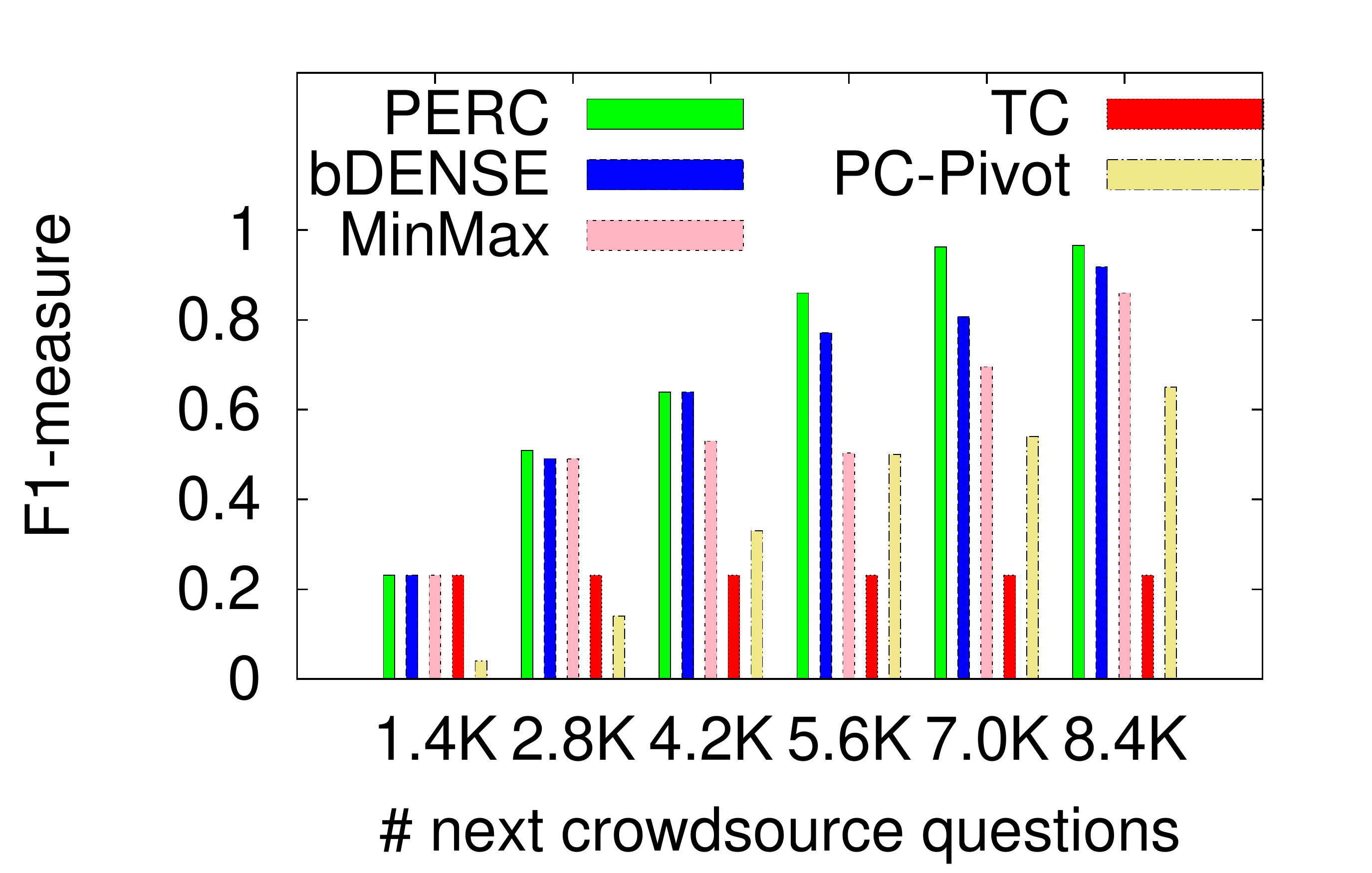}
\label{fig:f1_Landmarks}
}
\vspace{-5mm}
\caption{\small Accuracy improvement (F1-measure) for next crowdsourcing}
\label{fig:f1_next}
\vspace{-2mm}
\end{figure*}
\begin{figure*}[t!]
\vspace{-2mm}
\centering
\subfigure[\small {\em AllSports}] {
\includegraphics[scale=0.12, angle=270]{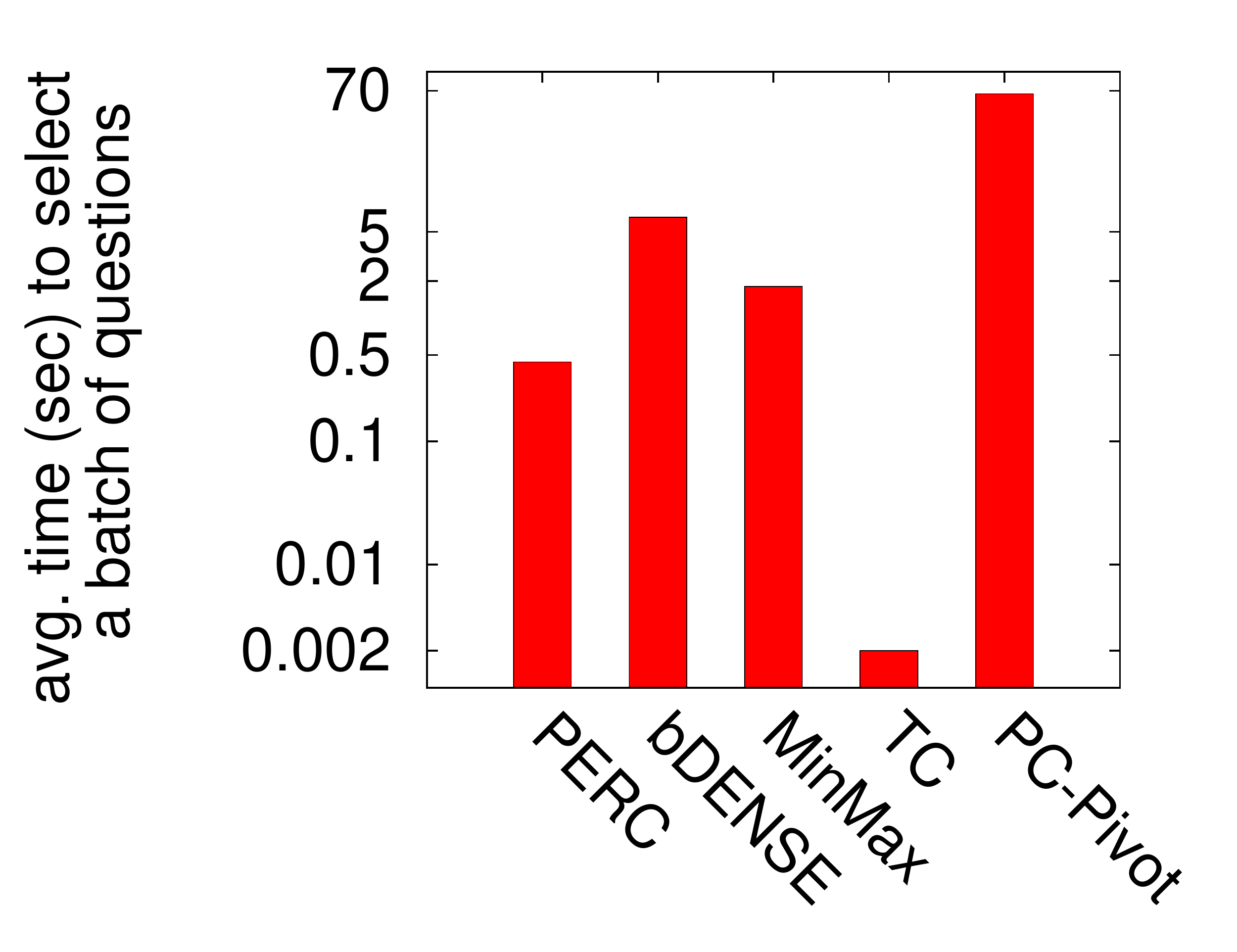}
\label{fig:time_allsports}
}
\subfigure[\small {\em Gymnastics}]  {
\includegraphics[scale=0.12, angle=270]{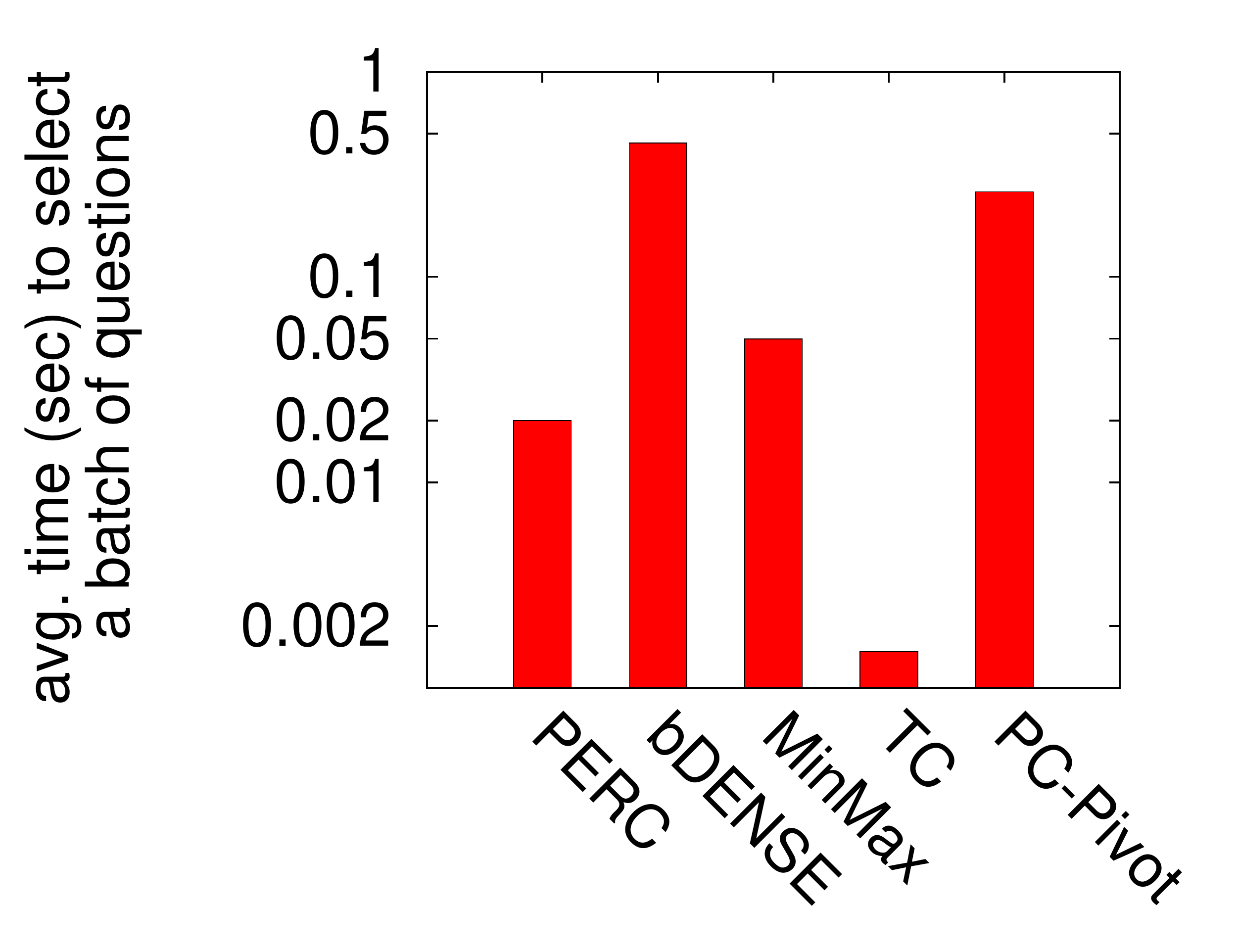}
\label{fig:time_gym}
}
\subfigure[\small {\em Landmarks}]  {
\includegraphics[scale=0.12, angle=270]{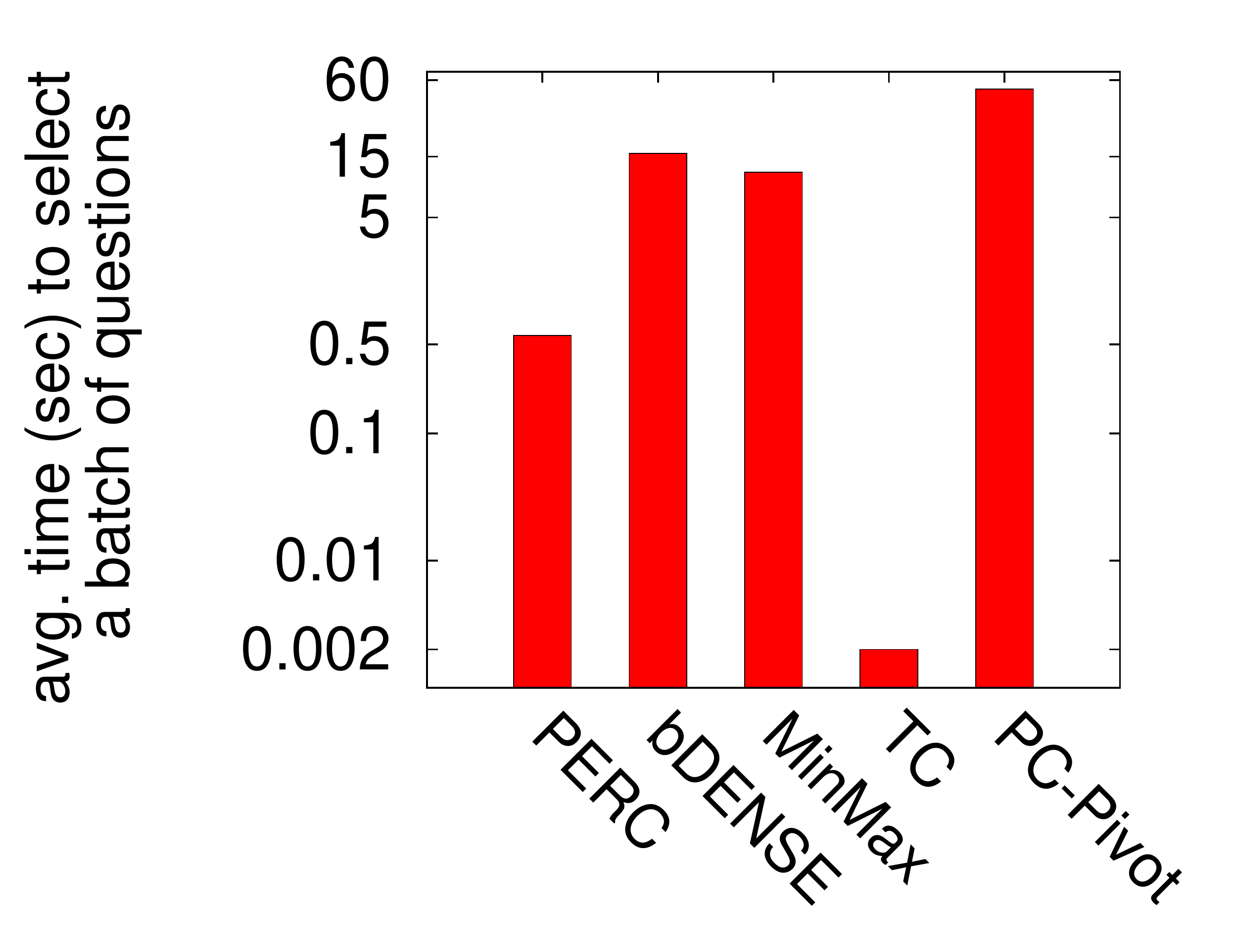}
\label{fig:time_Landmarks}
}
\vspace{-5mm}
\caption{\small Efficiency improvement: Computation time required to select a batch of next crowdsourcing questions}
\label{fig:time_next}
\vspace{-4mm}
\end{figure*}

\spara{$\bullet$ Evaluation Metrics.}

\vspace{-1mm}
\spara{Accuracy:}
After a certain number of answers are collected by the next crowdsourcing method, we apply ER algorithm for clustering.
To measure accuracy, we compare the output to the gold standard clustering. Specifically, we
employ precision (p) and recall (r), defined as follows.
\vspace{-1mm}
\begin{align}
& \displaystyle p = \frac{\# \text{record-pairs correctly reported as matching}}{\# \text{record-pairs reported as matching}} &
\end{align}
\vspace{-3mm}
\begin{align}
& \displaystyle r = \frac{\# \text{record-pairs correctly reported as matching}}{\# \text{matching record-pairs in gold clustering}} &
\end{align}
\vspace{-2mm}

Finally, we compute F1-measure, which is defined below.
\vspace{-2mm}
\begin{align}
\displaystyle F1\text{-measure}=2pr/(p+r)
\end{align}
Following previous works \cite{GKRW12,VG15}, we use F1-measure to demonstrate the accuracy of {\sf PERC} and other competitors.

\spara{Crowdsourcing Cost:} The crowdsourcing cost denotes the total number of distinct record pairs being crowdsourced.

\spara{Efficiency:} We report the average computation time required to select the next batch of crowdsourcing questions. Clearly,
this is the runtime of the algorithm to select the next batch of questions, and it excludes the crowdsourcing time (which would be
similar across different algorithms, for a given batch size).

\spara{$\bullet$ Compared Algorithms.}

\vspace{1mm}
\noindent {\bf Transitive Closure (TC):} This method selects, uniformly at random, one of those record pairs
for which the matching/ non-matching relationship cannot be inferred (via transitivity and anti-transitivity)
from the existing edges. Following \cite{WLG13,VBD14}, we consider majority voting while deciding on the next crowdsourcing results.
{\sf TC}-clustering never reaches F1-measure $\ge$ 0.75 over our datasets, which is because
this method does not consider conflicting evidences.

\vspace{1mm}
\noindent {\bf DENSE and bDENSE:}
{\sf DENSE} \cite{VG15} considers only either the set of positive edges,
or the set of negative edges between two disjoint record sets for calculating the strength of evidences,
denoted as the $\rho$-ratio (for details, see Introduction).
{\sf bDENSE} is a batch version of {\sf DENSE}, that selects multiple questions (having higher $\rho$-ratios)
to ask next, thereby allowing many crowd taskers to answer those questions in parallel. For ER, these methods apply SCC-clustering.

The authors in \cite{VG15} considered majority voting to decide on the next crowdsourcing results. Moreover, they
also assigned a fixed human accuracy of 0.9 (i.e., error rate = 0.1) on those answers.

\vspace{1mm}
\noindent {\bf MinMax:} For ER, \cite{GKRW12} finds all positive and
negative paths between a record pair. The weight of a path is determined by the smallest edge weight on that path.
Finally, the algorithm selects the maximum-weight path to decide whether the records are matching or not.
For next crowdsourcing, the authors proposed a hybrid strategy that prefers either a more certain matching pair,
or a less certain non-matching pair. As (1) {\sf MinMax} only considers the maximum-weight path
(and ignores all other paths) between a record pair for both ER and next crowdsourcing, and (2)
it does not consider the length of a path (intuitively, the error accumulated across a
short-length path would be less than that through a longer path), the method can easily produce less effective results.

\vspace{1mm}
\noindent {\bf PC-Pivot:} For ER, \cite{WXL15} uses pivot-based correlation clustering. The clustering refinement
phase consists of either splitting, where nodes are removed from clusters; or merging, where two clusters are combined.
The problem is that every node is considered individually, and edges connecting to that node are used to calculate the
respective benefit. Hence, the method may fail to capture the strength of the entire clustering, resulting in higher
crowdsourcing cost in order to achieve a reasonable ER accuracy.
\begin{figure*}[t!]
\centering
\subfigure[\small Cost]  {
\includegraphics[scale=0.13, angle=270]{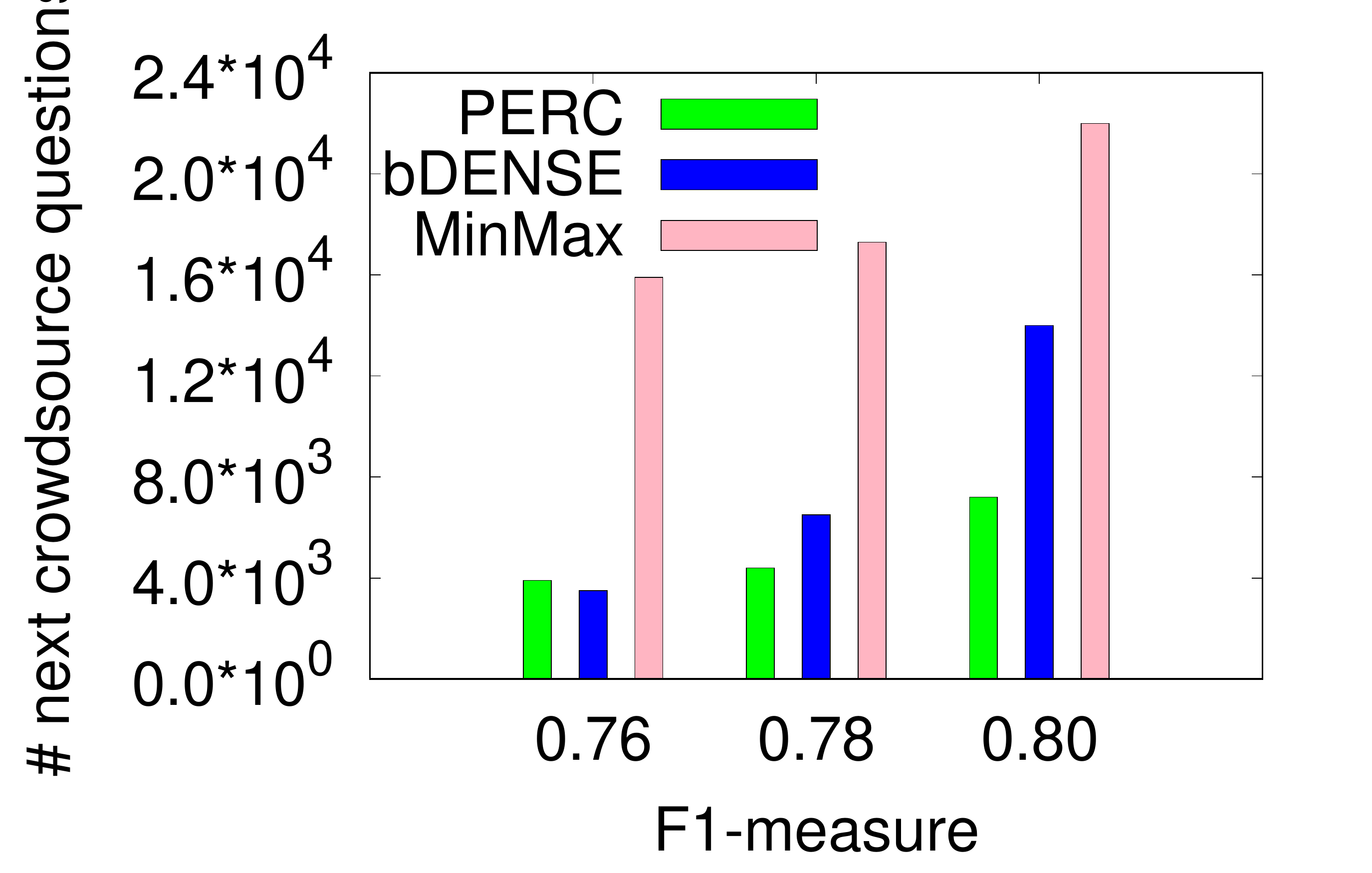}
\label{fig:cost_cora}
}
\vspace{-1mm}
\subfigure[\small Accuracy] {
\includegraphics[scale=0.13, angle=270]{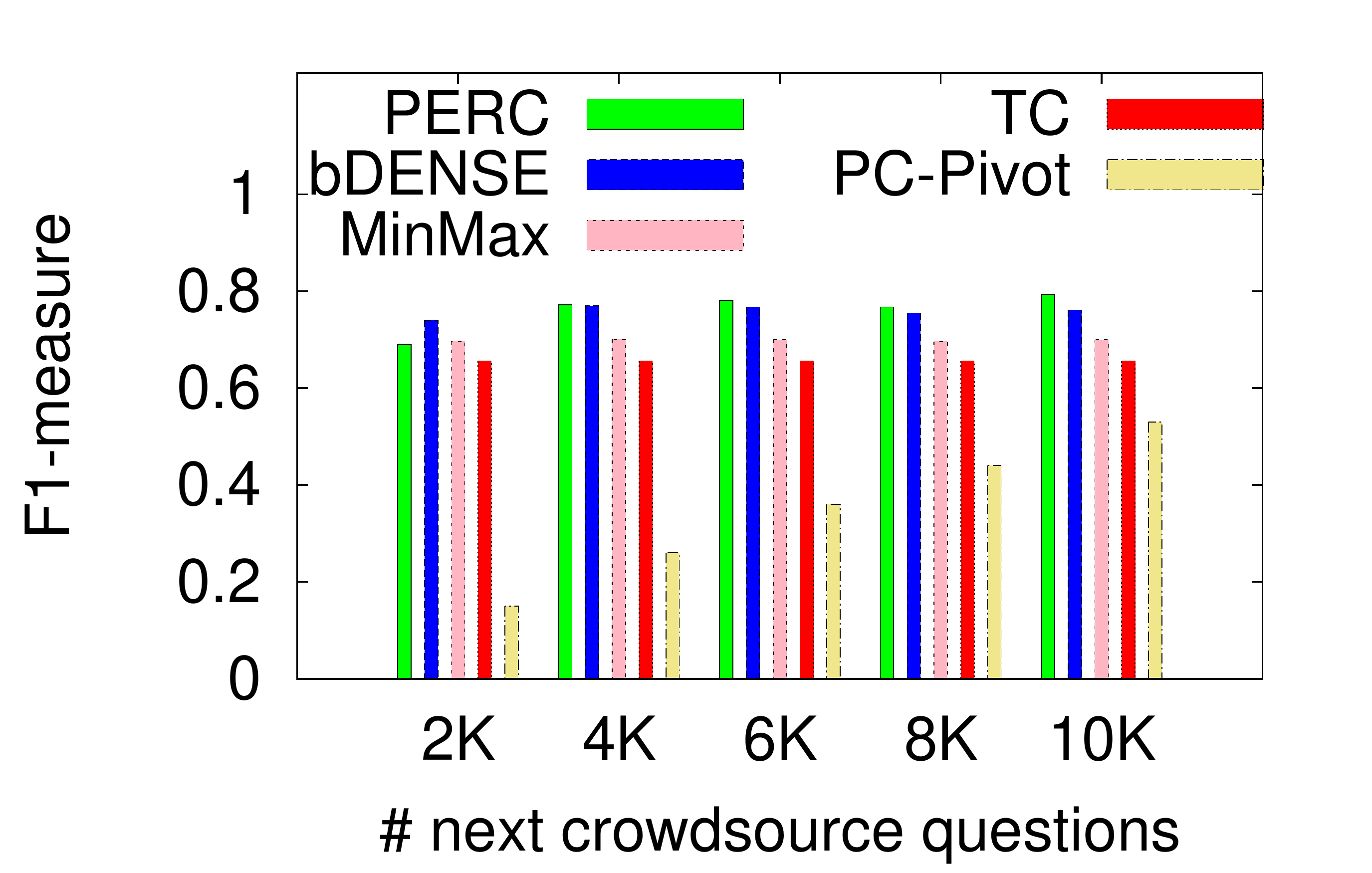}
\label{fig:f1_cora}
}
\vspace{-1mm}
\subfigure[\small Efficiency]  {
\includegraphics[scale=0.13, angle=270]{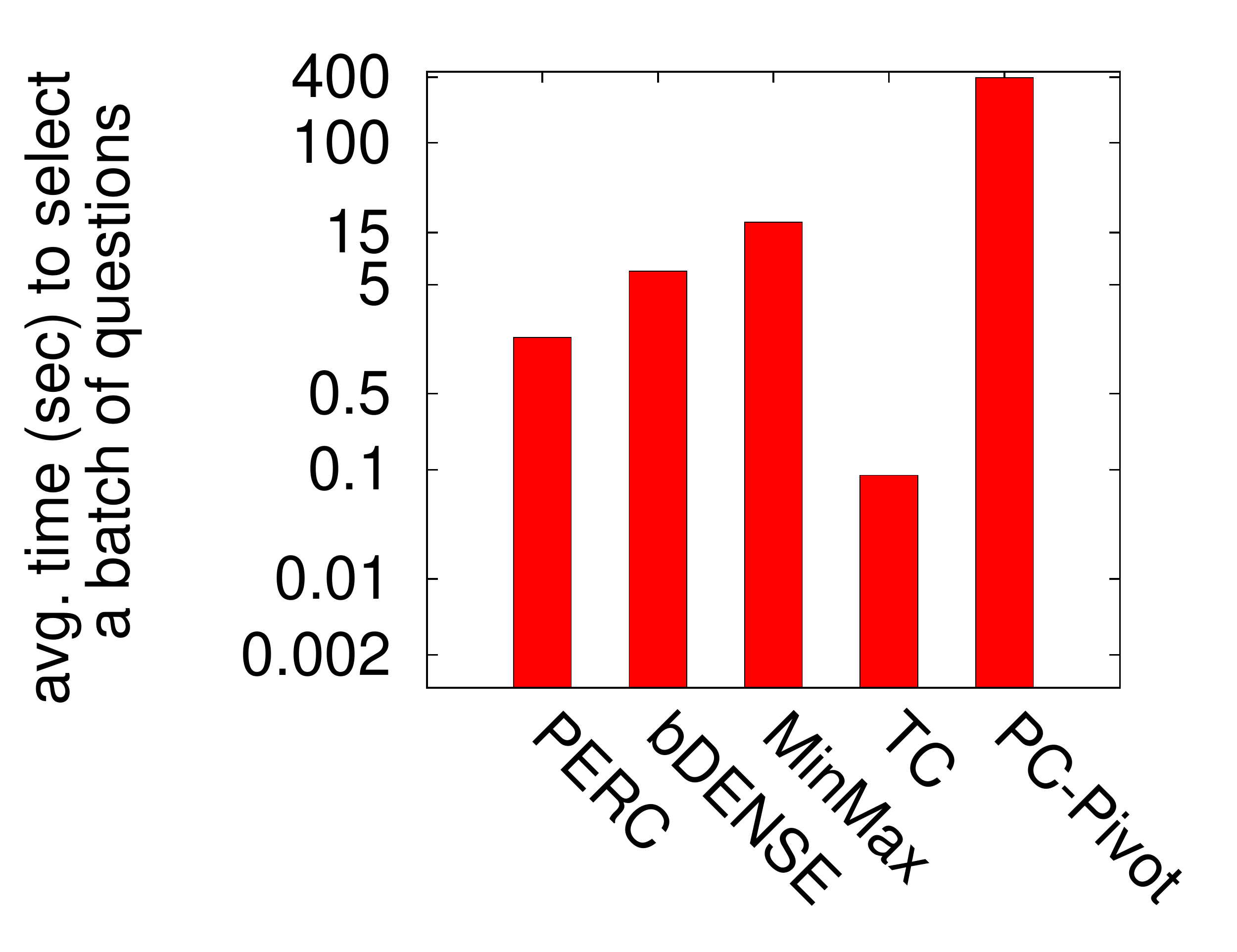}
\label{fig:time_cora}
}
\vspace{-5mm}
\caption{\small Accuracy, cost, and efficiency improvements over {\em Cora} dataset}
\label{fig:cora}
\vspace{-3mm}
\end{figure*}
\vspace{-1mm}
\subsection{Next Crowdsourcing Results}

We started with different
numbers of initial edges and batch sizes based on the size of our datasets. In particular, we had about 2K, 0.2K, and 1.4K initial
crowdsourced edges, respectively, for {\em AllSports}, {\em Gymnastics}, and {\em Landmarks} datasets. We set the batch size as
320, 40, and 120 questions, respectively, over these datasets.
\vspace{-1mm}
\subsubsection{Crowdsourcing Cost Improvement}

In Figure~\ref{fig:cost_next}, we show the number of next crowdsourcing questions required to reach a certain accuracy.
We consider F1-measure of 0.75 and above, because higher accuracy results are more important in real-world applications.
We do not show {\sf TC}-clustering, because it did not achieve an accuracy over 0.75 in all our datasets.
We find that the number of crowdsourcing questions required to obtain a higher accuracy is much less --- often by a margin
of 50\% --- for {\sf PERC}, in comparison to {\sf bDENSE}, {\sf PC-Pivot}, and {\sf MinMax}.
For example, to achieve F1-measure of 0.95 in the {\em Gymnastics} dataset, {\sf PERC},
{\sf bDENSE}, {\sf PC-Pivot}, and {\sf MinMax} require 863, 1792, 3360, and 1866 next crowdsourcing questions, respectively. These results demonstrate the effectiveness of {\sf PERC}
in reducing crowdsourcing cost.
\vspace{-1mm}
\subsubsection{Accuracy Improvement}

In Figure~\ref{fig:f1_next}, we illustrate accuracy improvements of {\sf PERC} over state-of-the-art approaches.
We observed that the F1-measure of {\sf PERC} increases at a higher rate and quickly reaches around 0.95 with less
number of next crowdsourcing questions, compared to other methods, in all our datasets. As an example, with about 12K next crowdsourcing questions over {\em AllSports},
the F1-measure of {\sf PERC} is 0.95, whereas for {\sf bDENSE}, {\sf MinMax}, {\sf PC-Pivot}, and {\sf TC}-clustering,
the F1-measures are 0.75, 0.79, 0.58, and 0.67,
respectively. These results demonstrate the accuracy improvements of {\sf PERC} next crowdsourcing algorithm.
\begin{figure}[tb!]
\centering
\includegraphics[scale=0.13, angle=270]{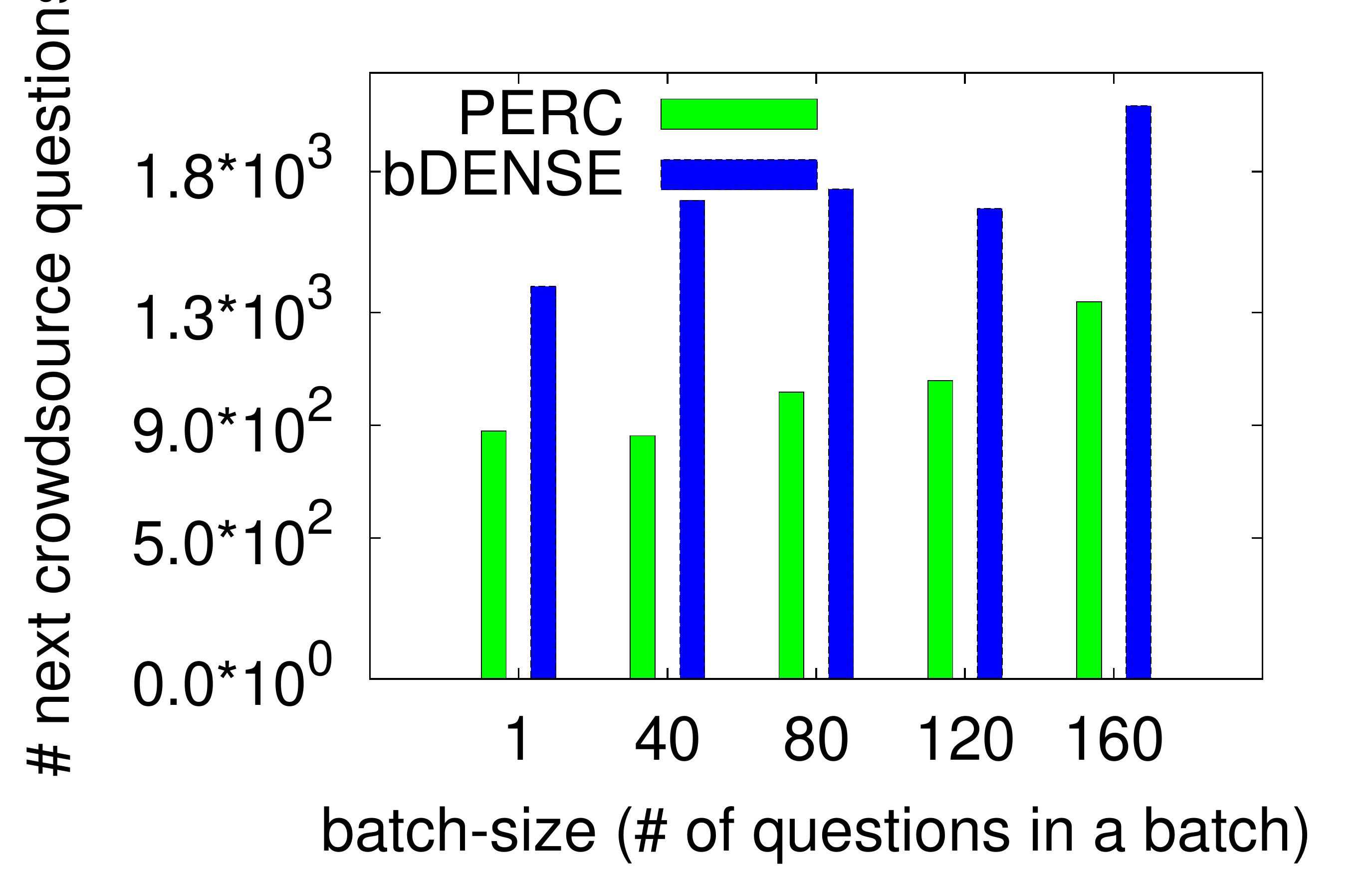}
\vspace{-4mm}
\caption{\small Varying batch-sizes: \# next crowdsourcing questions required to reach F1-measure=0.95, {\em Gymnastics}}
\label{fig:vary_batch}
\vspace{-6mm}
\end{figure}

\vspace{-1mm}
\subsubsection{Efficiency Improvement}

We compare the average computation time required to select a batch of next crowdsourcing questions, which is computed as follows.
We first measure the computation time to select all next crowdsourcing questions in order to reach a certain accuracy, e.g., F1-measure of 0.9 for {\sf PERC}, {\sf bDENSE}, {\sf PC-Pivot},
and {\sf MinMax}. One may recall that {\sf PERC} next crowdsourcing might trigger an update of the previous maximum-likelihood clustering.
We empirically found that these updates happen 20$\sim$25\% of the times after next crowdsourcing, and the times consumed for such
re-clusterings are also added in the total time required for {\sf PERC}. Since {\sf TC}-clustering does not achieve such a high accuracy,
we instead consider the time required to obtain the highest possible accuracy
via {\sf TC}-clustering. Next, we divide this time by the total number of batches issued to crowd workers, and report this value
as the average computation time to select one batch of next crowdsourcing questions for the respective methods.

\vspace{1mm}
Figure~\ref{fig:time_next} shows that the average time for one batch selection is at least an order of magnitude faster in case of {\sf PERC}, compared to that of {\sf bDENSE}, {\sf PC-Pivot}, and {\sf MinMax}.
We note that the Y-axis is logarithmic in these figures. For example, with the {\em Landmarks} dataset, the average time to select one batch (with 120 questions)
using {\sf PERC} is only 0.5 sec, whereas it requires about 15 sec, 12 sec, and 51 sec, respectively, to select a batch of same size using {\sf bDENSE}, {\sf MinMax}, and {\sf PC-Pivot}.
Thus, our empirical results illustrate that {\sf PERC} is at least an order of magnitude faster compared to {\sf bDENSE}, {\sf MinMax}, and {\sf PC-Pivot}, in terms of selecting the
next crowdsourcing questions.
\vspace{-1mm}
\subsubsection{Results with Cora Dataset}

We present next crowdsourcing results over the larger {\em Cora} dataset in Figure~\ref{fig:cora}.
We started with 2K initial crowdsourced edges, and we set the batch size as
300 questions. Figures~\ref{fig:cost_cora} and \ref{fig:f1_cora} demonstrate the cost and accuracy
improvements of {\sf PERC}. For example,
to achieve F1-measure = 0.8, {\sf PERC} requires about 7.2K questions, whereas {\sf bDENSE} and {\sf MinMax}
require around 14K and 22K questions, respectively.
The maximum F1-measure reached by {\sf PC-Pivot} over {\em Cora} is 0.74.
In Figure~\ref{fig:time_cora}, we compare the average computation times required to select a batch of 300 next crowdsourcing questions over {\em Cora}
dataset. The Y-axis is logarithmic. As earlier, {\sf PERC} is 5$\sim$15 times faster than both {\sf bDENSE}
and {\sf MinMax}, e.g., {\sf PERC} requires 1.5 sec to select a batch of 300 questions, whereas {\sf bDENSE} and {\sf MinMax}
consume 7 sec and 20 sec, respectively, for the same.
\vspace{-1mm}
\subsubsection{Varying Batch Sizes}

We analyze the impact of varying batch sizes on crowdsourcing cost and accuracy (Figure~\ref{fig:vary_batch}).
Smaller batch sizes help in improving the accuracy and to reduce the crowdsourcing cost.
This is because we do not know the corresponding edge probabilities apriori;
and hence, by issuing multiple questions in batches, the overall quality would decrease.
However, asking questions in batches reduces the {\em overall} running time (i.e., next batch selection time + crowdsourcing time),
since many crowd workers would be able to answer the questions in a batch in parallel.
In Figure~\ref{fig:vary_batch}, we show the number of next crowdsourcing questions
required to reach F1-measure=0.95 for {\sf PERC} and {\sf bDENSE}. We present our results
over {\em Gymnastics} dataset. As expected, this crowdsourcing cost decreases with smaller batch sizes,
for both these methods. We also observed that {\sf PERC} outperforms {\sf bDENSE} in terms of crowdsourcing cost
under all batch sizes.

\vspace{-2mm}
\section{Related Work}
\label{sec:related}


\spara{$\bullet$ Crowdsourcing in Data Management.} Recently, crowdsourcing has been adopted in
video and image annotations, search relevance, and natural language processing \cite{GWKP11,ARS08}.
Several systems have been developed to incorporate
human work into a database/mobile system, e.g., CrowdDB, Deco, CrowdSearch,
CDAS, CrowdForge, TurKit, and Qurk \cite{MWKMM11,LLOSWZ12}.
There are also studies on leveraging crowd's ability to
improve data management tasks, e.g., selection, sort,
skyline, join, mining, classification, and max/top-k \cite{MP15,CLM15}.

\spara{$\bullet$ Crowdsourced Entity Resolution (ER).} An important problem in crowdsourced
ER is to reduce the number of questions asked to workers, e.g., a clustering-based method \cite{WKFF12} where each question is a group
of records and asks workers to classify the records into different clusters.
Demartini et. al. \cite{DDM12} and Jeffrey et. al. \cite{JFH08} designed crowdsourcing systems based on a probabilistic framework,
but does not employ transitivity to reduce the
crowdsourcing cost. Wang et. al. \cite{WLKFF13} and Vesdapunt et. al. \cite{VBD14} utilized transitivity to reduce the number
of questions. Various models to select high-quality questions
were developed in \cite{WLG13,WJJ12}. The most recent work \cite{CLLDF16} used a partial order approach,
which additionally requires each entities having multiple attributes.
More importantly, all these works assume no crowd error, or employ majority voting.

Recently, {\sf MinMax}, {\sf PC-Pivot}, and {\sf DENSE} \cite{GKRW12,VG15,WXL15} directly incorporated
crowd errors in ER tasks. However, as we stated earlier, these methods consider ad-hoc, local features
to select next questions, such as individual paths, nodes, or the set of either positive or negative edges.
Hence, they generally fail to capture the strength of the entire clustering, resulting in higher crowdsourcing cost
in order to achieve a reasonable ER accuracy.

\spara{$\bullet$ Dealing with Crowdsourcing Errors.} Quality control is critical in crowdsourcing \cite{LLOSWZ12}.
Machine learning techniques have been employed to determine the quality of the crowd, e.g., \cite{KOS11,JGP13,DS09}.
Orthogonal to these works, our proposed solution incorporates crowd errors while performing next crowdsourcing and ER tasks.

\spara{$\bullet$ Entity Resolution Algorithms.} Entity resolution (ER), also known as entity reconciliation, deduplication, or
record linkage, is well studied in data cleaning and integration.
Many ER algorithms have been proposed based on different input settings, e.g.,
single-pass clustering, star clustering, cut clustering, correlation
clustering, and Markov clustering \cite{GM13,KSS06}.
We used correlation clustering because this is the most natural setting for clustering a set
of records that are connected by both positive and negative edges \cite{ES09}. Besides, our contribution
--- reliability-based next crowdsourcing question selection is orthogonal to the specific ER method employed.

\vspace{-1mm}
\section{Conclusions}
\label{sec:conclusions}

We studied crowdsourced entity resolution together with erroneous crowd answers.
Our solution {\sf PERC} does not require any user-defined threshold values, 
and no apriori information about the error rate of crowd workers. We formulated the 
problem considering an uncertain graph model and using
possible world semantics with edge independence. We employed the notion
of reliability in uncertain graphs to identify the most effective next
crowdsourcing questions.
Based on detailed empirical results with four real-world datasets, {\sf PERC}
improves the accuracy by 15\%, reduces the crowdsourcing cost by 50\%, and also
decreases the next question selection time by an order of magnitude
compared to state-of-the-art approaches.

\vspace{-1mm}
\section{Acknowledgement}

Research was supported by MOE Tier-1 M401020000 and NTU M4081678.
Any opinions, findings, and conclusions in this publication are those of the authors,
and do not necessarily reflect the views of the funding agencies.

{\scriptsize
\bibliographystyle{abbrv} 
\bibliography{ref}
}

\appendix
\underline{Limitation of DENSE \cite{VG15} with running example.}
The Dense considers only either
the set of positive edges (i.e., edges with majority YES votes), or the set of negative edges (i.e., edges having majority NO votes)
between two disjoint record sets for calculating the strength of evidences. A metric {$\rho$}-ratio is defined, which finds the lack of strong
evidences for clustering, and {\sf DENSE} selects a pair to crowdsource that has the maximum $\rho$-ratio.
In particular, $\rho$-ratio between sets $A$ and $B$ is calculated as follows.
\begin{align}
	& \frac{P'_{Y1} \times P'_{Y2}}{P_{Y1}\times P_{Y2}} \times \displaystyle \min\{\frac{P'_N}{P_N},\frac{P'_Y}{P_Y}\} &
	\label{eq:rho}
\end{align}

Here, $Y1$ is the set of positive edges between $A$ and $R\setminus B$, $Y2$ the set of positive edges between $B$ and $R\setminus B$, $N$ the set of negative edges across $A$ and $B$,
and $Y$ the set of positive edges across $A$ and $B$. The set of all records are denoted by $R$.
Let the probability for an edge $a \in \{ Y1 \bigcup Y2 \bigcup Y \bigcup N \}$ being correct be $p(a)$, then we compute:
\begin{align}
	& P_{Y1} = \prod_{a\in Y1}p(a); \quad P_{Y2} = \prod_{a\in Y2}p(a); \quad P_{Y} = \prod_{a\in Y}p(a); \nonumber & \\
    & P'_{Y1} = \prod_{a\in Y1}(1-p(a)); \quad P'_{Y2} = \prod_{a\in Y2}(1-p(a)); \quad P'_{Y} = \prod_{a\in Y}(1-p(a)); \nonumber & \\
	& P_{N} = \prod_{a\in N}p(a); \quad P'_{N} = \prod_{a\in N}(1-p(a)) &
\end{align}
\vspace{-2mm}

Since $\rho$-ratios between the clusters $\langle \mathbb{C}_1,\mathbb{C}_2 \rangle$ and $\langle \mathbb{C}_3,\mathbb{C}_4\rangle$
have the same value, which is due to the weaker negative
edges, i.e. $\frac{P'_N}{P_N} = \frac{0.3}{0.7}$,
{\sf DENSE} assumes that asking a question across $\langle \mathbb{C}_1,\mathbb{C}_2 \rangle$ or $\langle \mathbb{C}_3,\mathbb{C}_4 \rangle$ is equivalent.
However, in reality, asking a question between clusters $\mathbb{C}_3$ and $\mathbb{C}_4$ is more beneficial. 

\end{document}